\documentclass[aps,prx,amsmath,amssymb,reprint,superscriptaddress]{revtex4-2}

\usepackage{hyperref}
\hypersetup{
	colorlinks,
	linkcolor={rgb:red,2; blue,1},
	citecolor={blue!60!black},
	urlcolor={blue!60!black}
}
\usepackage{graphicx}
\usepackage{amsmath,amssymb,amsthm,bbm,bm}
\usepackage{xcolor}
\usepackage{txfonts}
\usepackage{braket}  

\usepackage{bbold}

\newcommand{\ii}{\ensuremath{\mathrm{i}}}
\newcommand{\ee}{\ensuremath{\mathrm{e}}}

\renewcommand{\a}{\ensuremath{\hat{a}}}
\newcommand{\adag}{\ensuremath{\hat{a}^\dagger}}
\renewcommand{\b}{\ensuremath{\hat{b}}}

\renewcommand{\c}{\ensuremath{\hat{c}}}
\newcommand{\cdag}{\ensuremath{\hat{c}^\dagger}}

\graphicspath{{./figs/}}

\begin{document}
	
\title{Robust nonequilibrium surface currents in the 3D Hofstadter model}
	
\author{Mark T. Mitchison}
\email{mark.mitchison@tcd.ie}
\affiliation{School of Physics, Trinity College Dublin, College Green, Dublin 2, Ireland}
\author{\'Angel Rivas }
\email{anrivas@ucm.es}
\author{Miguel A. Martin-Delgado}
\email{mardel@ucm.es}
\affiliation{Departamento de F\'{\i}sica Te\'orica, Facultad de Ciencias F\'isicas, Universidad Complutense, 28040 Madrid, Spain.}
\affiliation{CCS-Center for Computational Simulation, Campus de Montegancedo UPM, 28660 Boadilla del Monte, Madrid, Spain.}	
	
	\begin{abstract}
Genuinely two-dimensional robust crosscurrents --- which flow against the natural direction of heat flux --- have been missing since the discovery of their one-dimensional counterpart. We provide a setup to realize them on a cubic three-dimensional (3D) lattice hosting a Hofstadter model coupled to two heat baths with different temperatures. We show that these currents exhibit dissipative robustness: they are stable against the presence of impurities and tilting of the gauge field in certain nonequilibrium configurations. Moreover, we find protected boundary currents with genuinely 3D robustness, i.e.~they are only stable if tunnelling can occur in all three spatial directions. The model also presents generic surface currents, which are robust for both bosonic and fermionic systems. We identify the underlying qualitative mechanism responsible for the robustness of the surface currents and the crucial role played by certain discrete symmetries. 
	\end{abstract}
	\maketitle
	
\section{Introduction}
\label{sec:intro}

The fascinating world of exotic quantum phenomena in condensed matter physics has traditionally been associated with low-dimensional systems, i.e.~those existing in one and two dimensions \cite{auerbach1994interacting,gonzalez2008quantum,ezawa2008quantum}. This has been true not only for quantum phase transitions, but also for transport phenomena \cite{Beenakker1997,rammer2018quantum,nazarov2009quantum}.  The advent of topological quantum computing on the one hand~\cite{KITAEV2003,Bombin2006,Dennis2002,Bombin2007}, and topological insulators and superconductors on the other~\cite{QiZhang,Hasan,Ando2013}, provides the opportunity to find captivating new quantum properties in systems in three dimensions and even formally in higher dimensions~\cite{Bombin2007PRB, Qi2008}. A particularly interesting avenue of research in this direction concerns the emergence of such phenomena in open quantum systems far from equilibrium~\cite{Diehl2011,Budich2015,Iemini2016,Linzner2016,Rivas2017,Kawabata2019,Song2019,Shavit2020,Gau2020,Lieu2020,McGinley2020,Flynn2021}, since these may exhibit novel behaviour that cannot occur in closed systems.

\begin{figure}[t]
	\includegraphics[width=\linewidth]{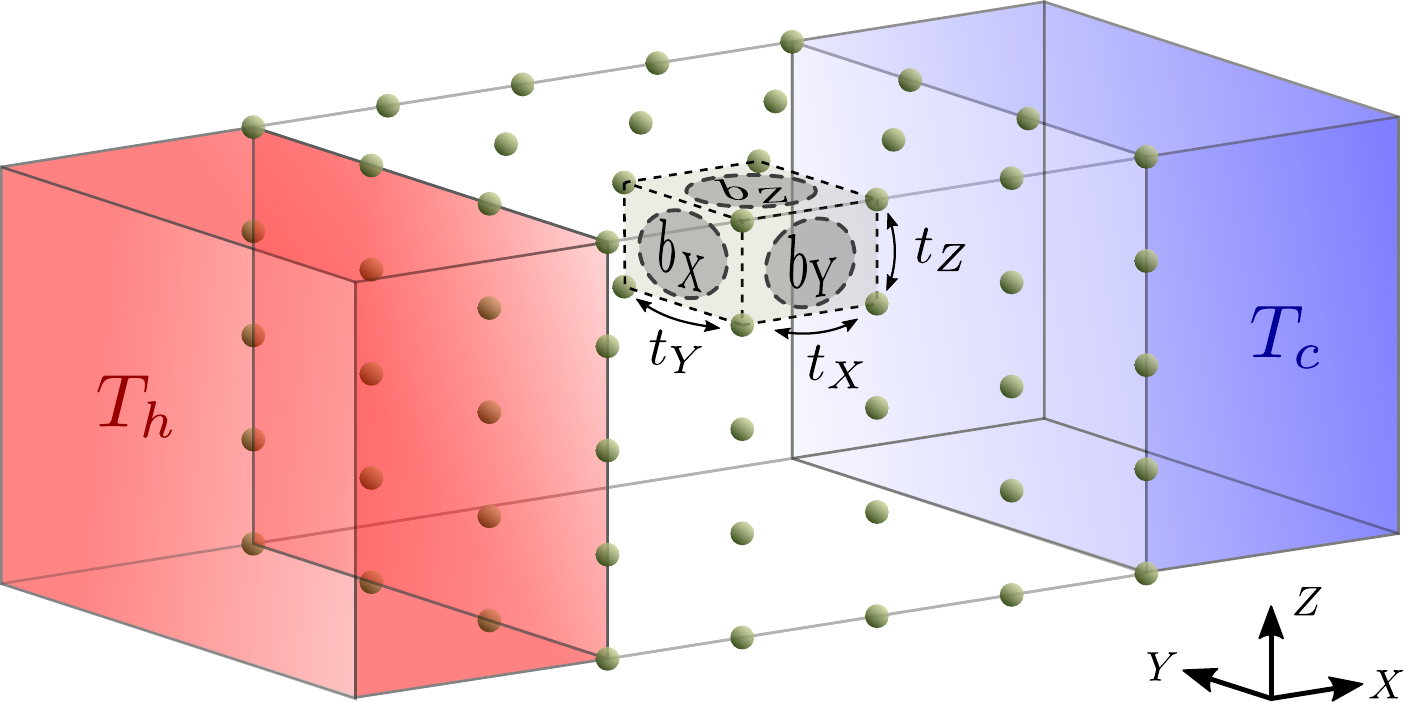}
\caption{Schematic of the 3D Hofstadter model coupled to hot and cold reservoirs, with respective temperatures $T_h>T_c $, along the $X$-axis. The coefficients $t_{X,Y,Z}$ indicate tunneling amplitudes along the three spatial directions and each face of the unit cell is crossed by the magnetic fluxes $b_X$, $b_Y$ and $b_Z$ depending on orientation of the gauge field.}
\label{fig:scheme}
\end{figure}

In this paper, we address an open problem in the study of transport in three-dimensional lattices that are out of thermodynamic equilibrium due to coupling to thermal baths at different temperatures. The subject of our study will be the current created by such a temperature gradient when the system has reached the steady state,  so that its entropy remains constant. For a two-dimensional (2D) Hofstadter lattice model of bosons, a one-dimensional edge current was recently found flowing in the opposite direction to the natural arrow of heat flow, and for this reason is called a crosscurrent~\cite{Rivas2017}. This exotic edge current, which has also been observed in other 2D models~\cite{Mitchison2021} (see Refs.~\cite{Guo2016,Guo2017,Xing2020} for quasi-one-dimensional studies), is robustly protected by symmetry properties with respect to the presence of point-like defects, and remains stable for a wide range of reservoir coupling strengths \cite{Mitchison2021}. Such crosscurrents were first thought to be an intrinsically one-dimensional phenomenon. Since then, the possibility of observing a truly two-dimensional crosscurrent that is stable in the presence of dissipation has remained open. Here, we fill this gap by constructing an explicit realization of a 2D crosscurrent as a boundary current of a three-dimensional (3D) Hofstadter lattice (see Fig. \ref{fig:scheme}). Moreover, we demonstrate the existence of robust surface currents that are stable with respect to defects in the presence of dissipation, regardless whether they are crosscurrents or not.

The boundary currents we consider here are descendants of the exotic surface physics that appears in 3D topological insulators \cite{QiZhang,Hasan,Ando2013,FuKaneMele3D2007,FuKane2008,HasanMoore3D,Moore3D,Roy3D,Hsieh2008,LOrtiz2016}, the main difference being that the latter are closed quantum fermionic systems as opposed to the open bosonic and fermionic systems that we consider. Typically, these 3D topological models can be divided into two generic categories. One hosts a layered structure of connected 2D topological nontrivial systems, in such a way that their exotic transport properties in 3D are inherited from the nontrivial 2D properties of each layer. However, there also exist systems with robust boundary states which are of genuinely 3D origin. For time-reversal invariant (TRI) models, these two classes correspond to the so-called ``weak'' and ``strong'' 3D topological insulators \cite{Hasan,FuKaneMele3D2007}, respectively. For the case of 3D lattices with broken time-reversal (TRB) symmetry, layered models are easy to construct, the 3D integer Hall effect being the simplest instance \cite{MontambauxQHE3D,Hasegawa3DQHE,KunsztQHE3D,KohmotoIntegerQHE3D,KoshinoButterfly3D,BernevigGraphite3D,TangQHE3D}. However, observing truly 3D robust boundary states for TRB systems is more challenging, and the most studied model, the so-called Hopf insulator \cite{Moore_Hopf,Deng_Hopf,Kennedy_Hopf}, presents topologically protected boundary modes but its topological features are unstable to the emergence of extra energy bands.  

By extending the ideas previously developed in Refs.~\cite{Rivas2017,Mitchison2021} for 2D topological systems coupled to thermal baths, in this work we show that it is indeed possible to induce genuinely 3D robust boundary transport in TRB systems. Namely, the robust surface currents we observe arise from a genuinely three-dimensional effect and are not the result of stacking two-dimensional models with corresponding one-dimensional robust edge currents. In this regard, we obtain the following notable results:
\begin{itemize}
\item[(i)] We find robustness of boundary currents, allowing for a surface crosscurrent, under perturbations created by the presence of a variety of defects in the 3D lattice, including point-, surface- and volume-like defects. 

\item[(ii)] This robustness is induced by the fulfillment of certain symmetries in such perturbations. 

\item[(iii)]  For particular gauge field orientations, there are 2D layered symmetries which can stabilize an unstable layered current (or vice versa), creating a 3D stabilization effect. 

\item[(iv)] Moreover, there is a genuine 3D symmetry independent of any field orientation, which protects the surface currents without any reference to their layered structure.
\end{itemize}

In the next section, we introduce our 3D lattice model coupled to thermal baths and detail the solution for its non-equilibrium steady state (NESS). Following Ref.~\cite{Mitchison2021}, we focus on the regime of weak system-bath coupling where the boundary currents become dominant. In this regime, the NESS is well approximated by the solution of a Lindblad master equation. In Sec.~\ref{sec:results}, we describe our results for the nonequilibrium current distribution obtained within this framework. Sec.~\ref{sec:conclusions} is devoted to our conclusions.
	
\section{Model}
\label{sec:model}

	
The Hofstadter model in three dimensions describes a cubic lattice of $L_X\times L_Y\times L_Z$ sites in the presence of a gauge field, governed by the Hamiltonian $\hat{H}_S = \hat{H}_0 + \hat{H}_X + \hat{H}_Y + \hat{H}_Z$, with ($\hbar=k_B=1$)
\begin{align}
\label{H0}
\hat{H}_0 &=  \sum_{x,y,z} \omega_0 \adag_{x,y,z} \a_{x,y,z}, \\
\hat{H}_X &=  \frac{-t_X}{2}\sum_{x,y,z}  \ee^{\ii 2\pi b_Z y}\adag_{x,y,z}  \a_{x+1,y,z} + \rm h.c., \\
\hat{H}_Y &=  \frac{-t_Y}{2}\sum_{x,y,z}  \ee^{\ii 2\pi b_X z }\adag_{x,y,z}  \a_{x,y+1,z} + \rm h.c.,\\
\hat{H}_Z &=  \frac{-t_Z}{2}\sum_{x,y,z}  \ee^{\ii 2\pi b_Y x}\adag_{x,y,z}  \a_{x,y,z+1} + \rm h.c.
\end{align}
Here, $\a_{x,y,z}^\dagger$ creates a particle on lattice site $(x,y,z)$, $t_{X,Y,Z}$ denote the hopping amplitudes in the three orthogonal directions, and $\mathbf{b} = (b_X,b_Y,b_Z)$ is an effective ``magnetic'' flux vector on the three faces of the unit cell, as illustrated in Fig. \ref{fig:scheme}, as a result of the presence of the gauge field.  Unless otherwise indicated, we set $t_X = t_Y = t_Z = t$. The on-site energy shift $\omega_0$ is chosen to be large enough to ensure that all single-particle eigenenergies are positive, but its value is otherwise immaterial: we take $\omega_0=10t$ in the following. 

The system lattice is coupled to independent reservoirs, hot at one end, $x=1$, and cold at the other end, $x=L_X$, with temperatures $T_h=\beta_h^{-1}$ and $T_c=\beta_c^{-1}$ and chemical potentials $\mu_h$ and $\mu_c$, respectively. We consider a system-reservoir interaction of the bilinear form
\begin{align}
\hat{H}_{SB}=\sum_{j\in \partial S}\sum_q g_{qj}(\a^\dagger_j \b_{qj}+\b_{qj}^\dagger \a_j),
\end{align} 
with $\partial S$ the boundary of the system, and $g_{qj}=0$ if $j\notin \{x=1\}\cup \{x=L_X\}$. Here, $\b_{qj}^\dagger$ creates a particle in mode $q$ of the reservoir coupled to the lattice site $j$. We assume that system and reservoir operators, $\a$ and $\b$, satisfy the same canonical commutation or anticommutation relations, depending on whether we consider bosonic or fermionic systems, respectively. The total Hamiltonian is given by 
\begin{align}
	\label{global_Hamiltonian}
\hat{H}=\hat{H}_S+\hat{H}_{SB}+\hat{H}_B,
\end{align}
with
\begin{align}
\hat{H}_B=\sum_{j\in \partial S} \sum_q \Omega_{q j}\b^\dagger_{qj} \b_{qj}.
\end{align}

For sufficiently long times the lattice system reaches a non-equilibrium steady state (NESS). Assuming that the coupling between system and reservoir is weak enough, and following the same steps as in \cite{Rivas2017} for the 2D case within the Born-Markov and secular approximations \cite{Breuer2002,RivasHuelga,GardinerZoller}, the NESS of the Hofstadter system can be well approximated by the stationary solution of a Gorini-Kossakowski-Lindblad-Sudarshan master equation  
\begin{equation}\label{Lindblad}
 \mathcal{L}\hat{\rho} =-\ii [\hat{H}, \hat{\rho}] + \mathcal{L}_h\hat{\rho} + \mathcal{L}_c\hat{\rho} = 0.
\end{equation}
The dissipators for the hot and cold baths are given respectively by
\begin{align}
	\label{hot_dissipator}
	\mathcal{L}_h = \gamma \sum_\alpha s_\alpha \left ( \bar{n}_h(\omega_\alpha)\mathcal{D}[\cdag_\alpha] + [1\mp \bar{n}_h(\omega_\alpha)]\mathcal{D}[\c_\alpha]  \right ),\\
		\label{cold_dissipator}
	\mathcal{L}_c = \gamma \sum_\alpha r_\alpha \left ( \bar{n}_c(\omega_\alpha)\mathcal{D}[\cdag_\alpha] + [1\mp \bar{n}_c(\omega_\alpha)]\mathcal{D}[\c_\alpha]  \right ),
\end{align}
with $\mathcal{D}[\hat{L}]\bullet  = \hat{L}\bullet \hat{L}^\dagger - \tfrac{1}{2}\{\hat{L}^\dagger \hat{L},\bullet\}$ and $\bar{n}_{h,c}(\omega)=[\ee^{\beta_{h,c}(\omega-\mu_{h,c})}\mp(-1)]^{-1}$, where the minus sign is for fermions and the plus sign is for bosons.  Furthermore,  $\c_\alpha=\sum_j U_{j\alpha}\a_j$ denotes a canonical ladder operator of the diagonalised 3D Hofstadter Hamiltonian, $\hat{H}_S=\sum_\alpha \omega_\alpha \c^\dagger_\alpha \c_\alpha$, where $U_{j\alpha}$ is the unitary matrix comprising the one-particle eigenvectors of $\hat{H}_S$. For simplicity, we have assumed that the reservoir spectral density is constant within the frequency range of interest (our results are independent of this choice). The coefficients $s_\alpha$ and $r_\alpha$ describe the dimensionless coupling strength of eigenmode $\alpha$ to the hot and cold baths, respectively:
\begin{align}\label{s_and_r}
s_\alpha=\sum_{y=1}^{L_Y}\sum_{z=1}^{L_Z}|U_{(1,y,z),\alpha}|^2,\quad 
	r_\alpha=\sum_{y=1}^{L_Y}\sum_{z=1}^{L_Z}|U_{(L_X,y,z),\alpha}|^2.
\end{align}

The current operators are defined via the continuity equation~\footnote{Here, the continuity equation is derived by considering the Heisenberg equations of motion generated by the microscopic Hamiltonian given in Eq.~\eqref{global_Hamiltonian}.} for the particle density $\hat{n}_{x,y,z}=\a^\dagger_{x,y,z} \a_{x,y,z}$ along the three spatial dimensions:
\begin{align}
\hat{J}^X_{x,y,z}&=\ii\frac{t_X}{2}\ee^{\ii 2\pi b_Z y} \a_{x,y,z}^\dagger \a_{x+1,y,z}+\mathrm{h.c.}\,, \nonumber\\
\hat{J}^Y_{x,y,z}&=\ii\frac{t_Y}{2}\ee^{\ii 2\pi b_X z} \a_{x,y,z}^\dagger \a_{x,y+1,z}+\mathrm{h.c.}\,, \nonumber\\
\hat{J}^Z_{x,y,z}&=\ii\frac{t_Z}{2}\ee^{\ii 2\pi b_Y x} \a_{x,y,z}^\dagger \a_{x,y,z+1}+\mathrm{h.c.},
\end{align}
so that
\begin{align}
\ii [\hat{H}_X,\hat{n}_{x,y,z}]&=\hat{J}^X_{x-1,y,z}-\hat{J}^X_{x,y,z}\,,\nonumber \\ 
\ii [\hat{H}_Y,\hat{n}_{x,y,z}]&=\hat{J}^Y_{x,y-1,z}-\hat{J}^Y_{x,y,z}\,, \nonumber\\
\ii [\hat{H}_Z,\hat{n}_{x,y,z}]&=\hat{J}^Z_{x,y,z-1}-\hat{J}^Z_{x,y,z}.
\end{align}
The mean currents can be obtained by computing the correlation matrix $\braket{\hat{c}^\dagger_\alpha \hat{c}_{\alpha'}}$, which can be done in a straightforward way by using the master equation~\eqref{Lindblad}.  The NESS correlation matrix takes the value \cite{Rivas2017,Mitchison2021}
\begin{align}\label{diagonal_NESS}
\braket{\hat{c}^\dagger_\alpha \hat{c}_{\alpha'}}=\delta_{\alpha\alpha'}\frac{s_\alpha \bar{n}_h(\omega_\alpha)+r_\alpha \bar{n}_c(\omega_\alpha)}{s_\alpha+r_\alpha},
\end{align}
so that only diagonal correlations in the eigenmode basis survive. In Ref.~\cite{Mitchison2021} it was shown that, for a 2D model, the solution of the Lindblad master equation gives an excellent approximation to the current distribution in the weak-coupling regime. Similarly, therefore, here we expect Eq.~\eqref{diagonal_NESS} to yield accurate values for the currents so long as $t_{X,Y,Z} \gg \gamma$. Our approximations are also valid if one of them vanishes, say $t_{Z} =0$, so long as the other two satisfy this condition, $t_{X,Y} \gg \gamma$.

	\begin{figure*}[t]
		\centering	\begin{minipage}{0.35\linewidth}\centering
		\includegraphics[width=\linewidth]{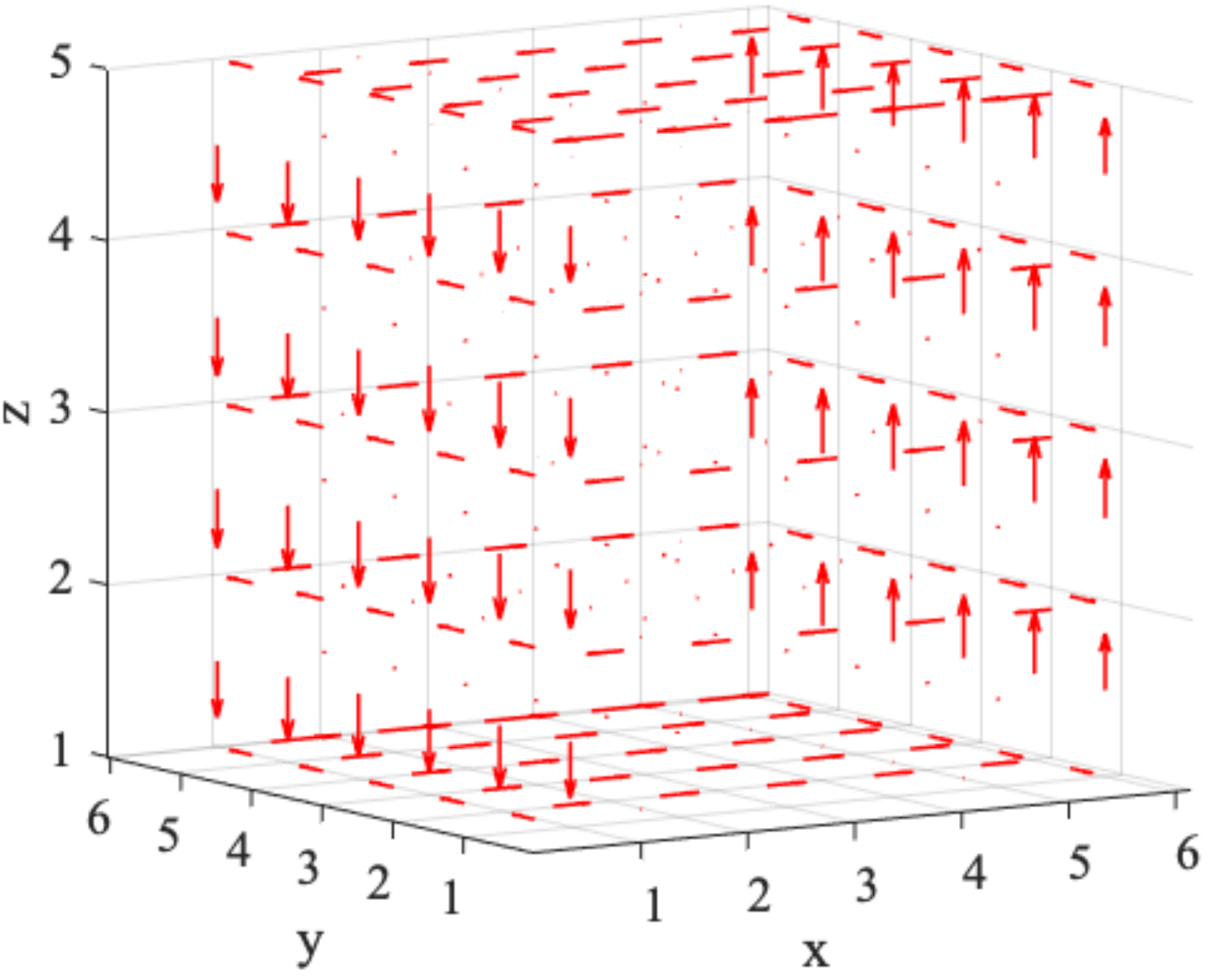}
	\end{minipage}
\hspace{10mm}
	\begin{minipage}{0.35\linewidth}\centering
		\includegraphics[width=\linewidth]{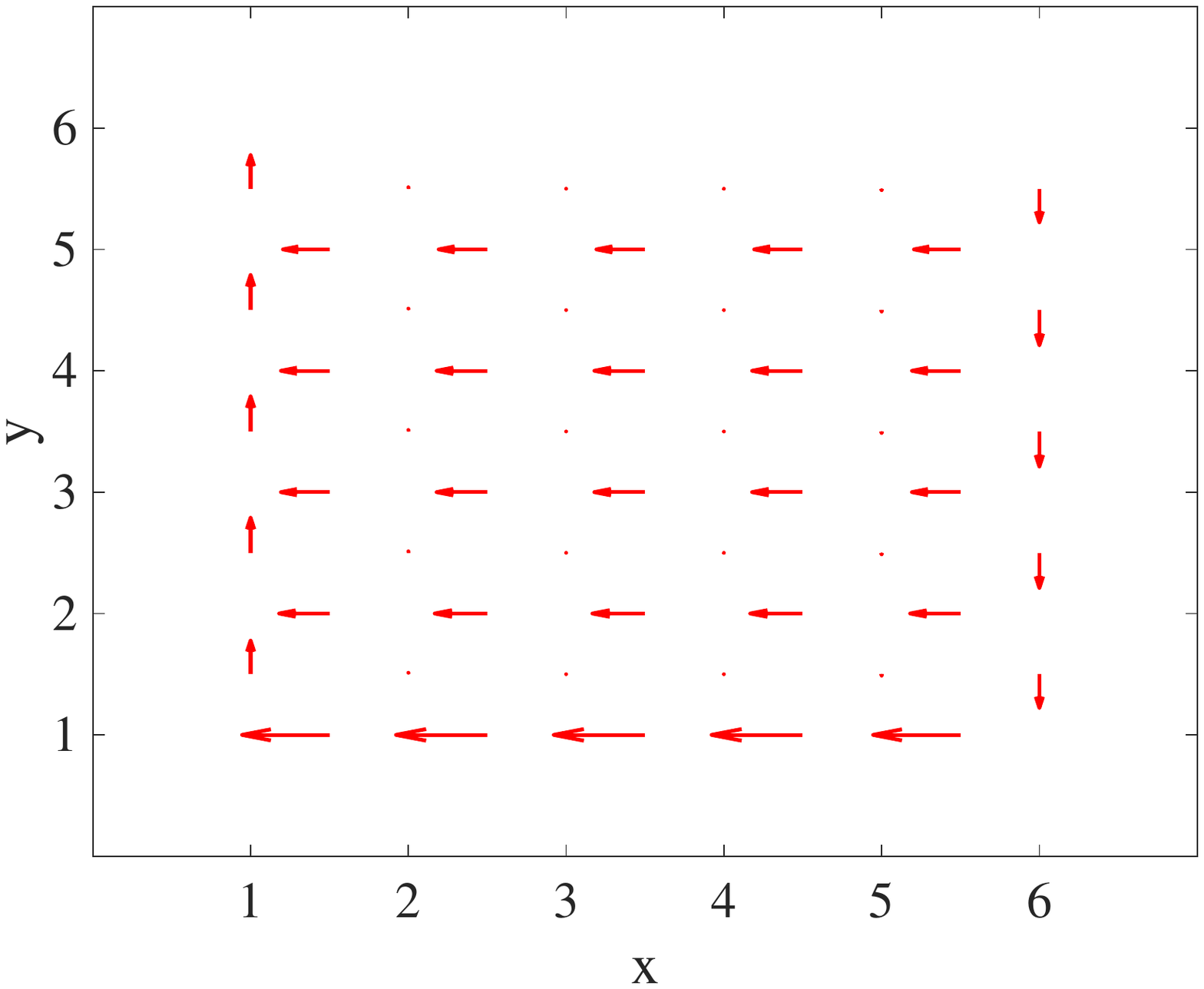}
	\end{minipage}
	\caption{Bosonic current distribution for a ``tilted'' field configuration with $\mathbf{b} = (0,0.1,0.1)$. Arrows indicate the direction of the current, with the length of the arrows proportional to the magnitude of the current. The length of the arrows is scaled relative to the largest current value within each plot. The right panel shows the topmost plane, demonstrating the existence of 2D cross-currents flowing from the hot reservoir ($x=1$) to the cold reservoir ($x=L_X$).}
	\label{fig:nodefects}
\end{figure*}

\section{Results}
	\label{sec:results}

\subsection{Symmetries and boundary currents}
\label{sec:symmetries}

A perfect 3D Hofstadter lattice, without any defect,  presents several symmetries depending on the orientation of the magnetic flux vector $\mathbf{b}$. We will be particularly interested in symmetries of $\hat{H}$ which interchange the hot and cold baths relative locations, as they have a nontrivial action on the NESS.  For instance,  if $\mathbf{b}=(0,0,b_z)$, the Hamiltonian is invariant under the symmetries $\hat{\Theta}\hat{\Sigma}_{yz}$,  $\hat{\Theta}\hat{R}_y(\pi)$ and $\hat{\Theta}\hat{R}_z(\pi)$.  Here, $\hat{\Theta}$ is the time-reversal operation (the same as complex conjugation,  in this case),  $\hat{\Sigma}_{yz}$ is a spatial reflection across the $yz$-plane cutting the center of the lattice, $\hat{\Sigma}_{yz}\a_{x,y,z}\hat{\Sigma}_{yz}^\dagger=\a_{L_x+1-x,y,z}$, and $\hat{R}_{y,z}(\pi)$ are $\pi$-rotations about the $x$ and $y$ axes, respectively, from the center of the lattice, $\hat{R}_y(\pi)\a_{x,y,z}\hat{R}_y(\pi)^\dagger=\a_{L_x+1-x,y,L_z+1-z}$,  and $\hat{R}_z(\pi)\a_{x,y,z}\hat{R}_z(\pi)^\dagger=\a_{L_x+1-x,L_y+1-y,z}$.  These are essentially 2D symmetries embedded in the 3D lattice, and they arise for particular orientations of $\mathbf{b}$ or particular values of $t_{X,Y,Z}$ but not for arbitrary directions of the magnetic field.  Nevertheless, for a general $\mathbf{b}$,  the Hamiltonian is always invariant under $\hat{\Theta}\hat{I}$, where $\hat{I}$ is a spatial inversion transformation with respect to the center of the lattice $\hat{I}\a_{\mathbf{r}}\hat{I}^\dagger=\a_{\mathbf{R}-\mathbf{r}}$, with $\mathbf{r}=(x,y,z)$ and $\mathbf{R}=(L_x+1,L_y+1,L_z+1)$.

The master equation is also invariant under these symmetries, because the Liouvillian~\eqref{Lindblad} inherits them from the Hamiltonian within the Born-Markov-secular (BMS) approximation. Of course, the total Liouvillian is not invariant because it describes the evolution of an open system and these symmetries involve the time-reversal operation, $\Theta$. Therefore, only a kind of weak symmetry can be expected, in the sense of Ref.~\cite{Buca_2012}. Each of the aforementioned symmetries, $\hat{S}$, is a weak symmetry~\cite{Buca_2012} for both the Hamiltonian part $\mathcal{L}_H(\rho) := [H,\rho]$ and dissipative part  $\mathcal{L}_D(\rho):= \mathcal{L}(\rho)+\ii\mathcal{L}_H(\rho)$ of the Liouvillian: namely, $\hat{S}\mathcal{L}_{H,D}(\rho) \hat{S}^{-1}=\mathcal{L}_{H,D}(\hat{S}\rho\hat{S}^{-1})$. Note, however, that the total Liouvillian $\mathcal{L} = -\ii \mathcal{L}_H + \mathcal{L}_D$ is not invariant: since $\hat{S}$ is anti-unitary, the Hamiltonian contribution flips sign. Nevertheless, since $[\mathcal{L}_H,\mathcal{L}_D]=0$ under the BMS approximation, the NESS is annihilated by both $\mathcal{L}_H$ and $\mathcal{L}_D$ individually. Therefore, a weak symmetry $\hat{S}$ of $\mathcal{L}_H$ and $\mathcal{L}_D$ is sufficient to ensure that the NESS is also invariant under $\hat{S}$.

Since the aforementioned weak symmetries interchange the reservoirs, the invariance of $\hat{H}$ also ensures that the dimensionless couplings~\eqref{s_and_r} obey $s_\alpha=r_\alpha$. As a result, the mode occupations given by Eq.~\eqref{diagonal_NESS} become independent of $s_\alpha$ and $r_\alpha$, being given simply by the average of the reservoir distribution functions:
\begin{equation}\label{mode_occ_symmetry}
	n(\omega_\alpha) = \frac{1}{2}\left [\bar{n}_c(\omega_\alpha) + \bar{n}_h(\omega_\alpha) \right ].
\end{equation}
This is a function of the frequency $\omega_\alpha$ only,  and leads to a steady-state correlation matrix that is independent of the spatial orientation of the reservoirs.  In such a case,  the induced current is highly suppressed in the bulk and flows essentially on the boundary of the 3D lattice.  To understand this, let us resort to a semi-classical picture \cite{Niu,Xiao2010,Cooper}.

Under periodic boundary conditions,  the 3D Hofstadter Hamiltonian can be written in the form
\begin{equation}
\hat{H}=\sum_{\mathbf{k}\in \mathrm{MBZ}} \mathbf{a}_{\mathbf{k}}^\dagger\cdot \mathbf{H}(\mathbf{k})\cdot \mathbf{a}_{\mathbf{k}},
\end{equation}
with $\mathbf{k}=(k_X,k_Y,k_Z)$ the quasimomentum, ranging inside the magnetic Bruillouin zone (MBZ), $ \mathbf{a}_{\mathbf{k}}^\dagger=(\hat{a}_{1\mathbf{k}}^\dagger,\hat{a}_{2\mathbf{k}}^\dagger,\ldots)$, and a hermitian matrix $\mathbf{H}(\mathbf{k})$ whose dimension and structure depend on $\mathbf{b}$.  For instance, for rational fluxes $\mathbf{b}=(p_X/q_X,p_Y/q_Y,p_Z/q_Z)$,  the number of eigenvalues $\omega_\alpha(\mathbf{k})$ of $\mathbf{H}(\mathbf{k})$ (energy bands) is given by the lowest common denominator
of the three fractions \cite{MontambauxQHE3D,Hasegawa3DQHE,KunsztQHE3D,KohmotoIntegerQHE3D}.

The action of a force $\mathbf{F}$ modifies the semiclassical equation for the velocity $\mathbf{v}_\alpha$ of a Bloch wave-packet in the band $\omega_{\alpha}(\mathbf{k})$  as 
\begin{equation}\label{Vsem}
\mathbf{v}_\alpha=\frac{\partial \omega_\alpha(\mathbf{k})}{\partial\mathbf{k}}-\mathbf{F}\times \bm{\mathcal{F}}\!\!_\alpha(\mathbf{k}).
\end{equation}
Here, $\bm{\mathcal{F}}\!\!_\alpha(\mathbf{k})=\bm{\nabla}\times \bm{\mathcal{A}}_\alpha$  is the Berry curvature vector, defined in terms of the Berry connection,  $\bm{\mathcal{A}}_\alpha(\mathbf{k})= \ii \langle u_\alpha (\mathbf{k})| \bm{\nabla}_{\mathbf{k}} |u_\alpha (\mathbf{k})\rangle $,  with $\mathbf{H}(\mathbf{k})|u_\alpha (\mathbf{k})\rangle=\omega_\alpha(\mathbf{k})|u_\alpha (\mathbf{k})\rangle$.

Under these conditions, we can generate a boundary by introducing a confining potential $V(\mathbf{r})$ in the semiclassical equation~\eqref{Vsem} for $\mathbf{v}_\alpha$.  Assuming that the potential varies slowly on the scale of the lattice, the energy of the wavepacket in the semi-classical approximation is simply $\omega_\alpha(\mathbf{k}) + V(\mathbf{r})$~\cite{Xiao2010}. Moreover, the potential induces a confining force $\mathbf{F}=-\bm{\nabla}V(\mathbf{r})$,  which approximately vanishes in the bulk and becomes very large near the edge of the confined region, so that
\begin{align}
	\label{edge-bulk-velocity}
\mathbf{v}_\alpha^\mathrm{(bulk)}\simeq  \frac{\partial \omega_\alpha(\mathbf{k})}{\partial\mathbf{k}}, \quad \mathbf{v}_\alpha^{\rm (edge)}\simeq \bm{\nabla}V(\mathbf{r})\times \bm{\mathcal{F}}\!\!_\alpha(\mathbf{k}).
\end{align} 
Since $\mathbf{F}$ is normal and points inward from the confining boundary,  a circulating current is induced along this edge with direction given by the orientation of the Berry curvature vector, $\bm{\mathcal{F}}\!\!_\alpha(\mathbf{k})=\bm{\nabla}\times \bm{\mathcal{A}}_\alpha$, which depends on the band $\alpha$, the direction of $\mathbf{b}$, and the specific value of $\mathbf{k}$. 

Thus, if there is an  invariance under any of the symmetries which guarantee that the NESS occupation number $n$ only depends on energy,  we may estimate the current density by \cite{Xiao2010}
\begin{align}
	\label{semiclassical_current}
\mathbf{I}(\mathbf{r})=\sum_{\alpha}\int_{\mathrm{MBZ}}d^3\mathbf{k}\, n[\omega_\alpha(\mathbf{k})+V(\mathbf{r})] \mathbf{v}_\alpha(\mathbf{r},\mathbf{k}).
\end{align}
Now,  inside the confined region $V(\mathbf{r})\ll \omega_\alpha(\mathbf{k})$,  and 
\begin{align}
	\mathbf{I}(\mathbf{r})&\simeq\sum_{\alpha}\int_{\mathrm{M BZ}}d^3\mathbf{k}\, n[\omega_\alpha(\mathbf{k})] \left[\mathbf{v}_\alpha^\mathrm{(bulk)}+\mathbf{v}_\alpha^{\rm (edge)}\right].
\end{align}
Since there is no privileged direction of $\mathbf{v}_\alpha^{\mathrm{(bulk)}}$ on each energy shell (because of the invariance under the aforementioned symmetries), $\omega_\alpha(\mathbf{k})=\omega_\alpha(-\mathbf{k})$ and hence $\frac{\partial \omega_\alpha(\mathbf{k})}{\partial\mathbf{k}}=-\frac{\partial\omega_\alpha(-\mathbf{k})}{\partial\mathbf{k}}$. Therefore, the contribution of $\mathbf{v}_\alpha^{(\rm bulk)}$ to the integrand is odd and integrates to zero. This cancellation of bulk currents is referred to as the \textit{erasure effect}~\cite{Rivas2017,Mitchison2021}. The remaining contribution to the current close to the edge is thus given by
\begin{align}
	\label{semi-classical_edge_current}
	\mathbf{I}(\mathbf{r})&\simeq\sum_{\alpha}\int_{\mathrm{M. B.Z.}}d^3\mathbf{k}\,n[\omega_\alpha(\mathbf{k})]\mathbf{v}_\alpha^{\rm (edge)}.
\end{align}
The monotonically decreasing behaviour of  $n(\omega)$ as a function of energy [Eq.~\eqref{mode_occ_symmetry}] creates an imbalance of the contributions to $\mathbf{I}(\mathbf{r})$ within the Brillouin zone in regions where $\mathbf{v}_\alpha^{\rm (edge)} \propto \nabla V(\mathbf{r}) \neq 0$. Thus, Eq.~\eqref{semi-classical_edge_current} predicts a net circulating surface current.   As a result, any magnetic flux vector $\mathbf{b}$ non-parallel to the temperature gradient induces a surface crosscurrent.  

Fig.~\ref{fig:nodefects} illustrates the surface crosscurrents for a cuboid geometry in the bosonic case.  Here, and in all the following examples, we take the reservoir temperatures to be $T_h=t$ and $T_c=0.01t$, and for bosons always set $\mu_\alpha=0$. For visual clarity, we display relatively small systems but large enough to avoid finite-size effects. Qualitatively similar results are obtained for other temperature values and larger lattices. The left panel of Fig.~\ref{fig:nodefects} demonstrates that all currents are restricted to the boundary of the system and vanish in the bulk.  The right panel of Fig.~\ref{fig:nodefects} zooms in on the topmost boundary of the system, demonstrating that particles and energy flow \textit{against} the temperature gradient within a two-dimensional surface. The occurrence of such two-dimensional crosscurrents represents the first main result of this work.

\subsection{Robustness under defects}

\begin{figure*}[t]
	\includegraphics[width=\linewidth, trim = 0mm 60mm 0mm 5mm, clip]{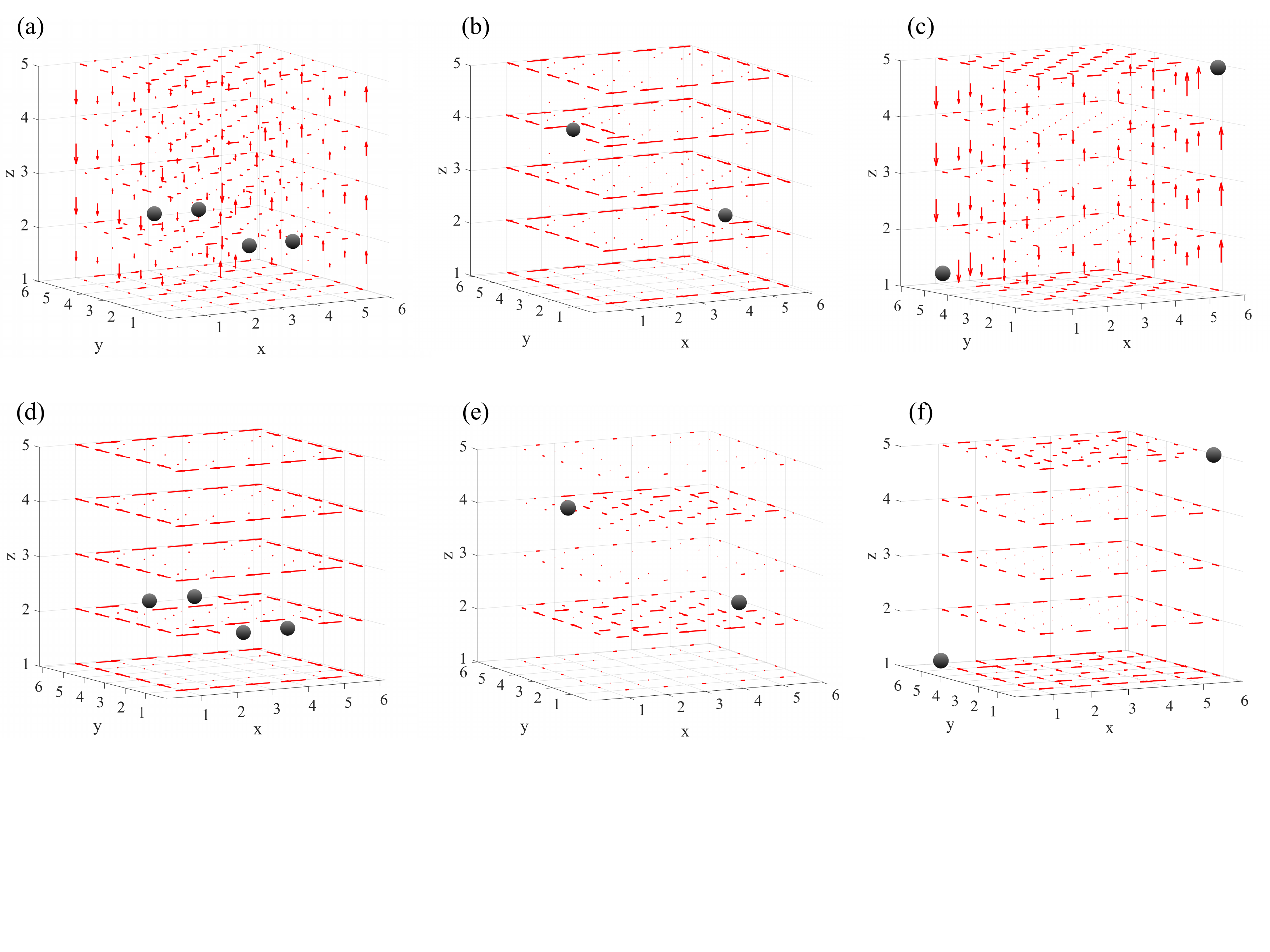}
	\caption{Robustness of bosonic boundary currents to defects  (gray circles), (a--c) in the three-dimensional case with $t_X = t_Y = t_Z= t $ and  (d--f) in the two-dimensional case with $t_Z=0$ and $t_X=t_Y = t$. (a,d) Tilted magnetic field, $\mathbf{b} = (0.1,0.1,0.1)$, with defects symmetric under the reflection $\hat{\Sigma}_{xy}$. (b,e) Vertical magnetic field, $\mathbf{b} = (0,0,0.1)$, with defects symmetric under the rotation $\hat{R}_{y}(\pi)$. (c,f) Tilted magnetic field, $\mathbf{b} = (0.1,0.1,0.1)$, with defects symmetric under the lattice inversion $\hat{I}$. \label{fig:defects} }
\end{figure*}

The erasure effect of the bulk currents sustains the appearance of the chiral current in the symmetric situation, but it is spoilt in the nonsymmetric case because the occupation number is no longer only a function of energy alone.  This situation arises when the lattice is not perfect, e.g. due to the presence of defects.  A defect is modeled here by a point-like impurity that induces a large on-site energy shift, i.e.,~a term $\Delta {\a}^\dagger_{\mathbf{r}}{\a}_{\mathbf{r}}$ added to the Hamitonian, where $\Delta \gg \omega_0, t$ and $\mathbf{r}$ is the position of the impurity.  Nevertheless, if the defect configuration complies with any of the protecting symmetries, such that \eqref{mode_occ_symmetry} remains valid, the surface current will be stable.

For instance, in the case $t_Z=0$, the system reduces to a layered stack of independent 2D Hofstadter lattices. The surface current is robust provided that the defect configuration satisfies either of the two symmetries $\hat{\Theta}\hat{\Sigma}
_{yz}$ or $\hat{\Theta} \hat{R}_z(\pi)$~\cite{Rivas2017}.  This is so independently of the orientation of $\mathbf{b}$, as shown in Fig.~\ref{fig:defects}(d). However, if $\mathbf{b}$ is not orthogonal to the temperature gradient, any nonzero $t_Z$ spoils the stability of currents 2D protected by $\hat{\Theta} \hat{\Sigma}
_{yz}$, as this operation ceases to be a symmetry of the 3D Hofstadter Hamiltonian. This is illustrated in Fig.~\ref{fig:defects}(a), where a chaotic current pattern throughout the edges and bulk of the system is induced by the defects. The 2D symmetry $\hat{\Theta} \hat{R}_z(\pi)$ is even more fragile as any $\mathbf{b}$ not parallel to some lattice vector destabilizes the boundary currents in 3D.

Remarkably, we also find the converse situation: starting from a stack of independent 2D Hofstadter layers with unstable currents, we can make them stable by switching on $t_Z$. This is shown in Figs.~\ref{fig:defects}(b,e), where there are two defects connected by the rotation $\hat{R}_y(\pi)$. Since $\hat{\Theta}\hat{R}_y(\pi)$ is not a symmetry of the model for $t_Z=0$, the defects destroy the edge currents [Fig.~\ref{fig:defects}(e)]. However, for $\mathbf{b}$ aligned to the $Z$ direction, $\hat{\Theta} \hat{R}_y(\pi)$ becomes a symmetry for $t_Z\neq 0$, and this produces a genuinely 3D stabilization effect [Fig.~\ref{fig:defects}(e)].

\subsection{Robustness under defects and arbitrary magnetic tilting}

\begin{figure}[b]
	\includegraphics[width=0.7\linewidth]{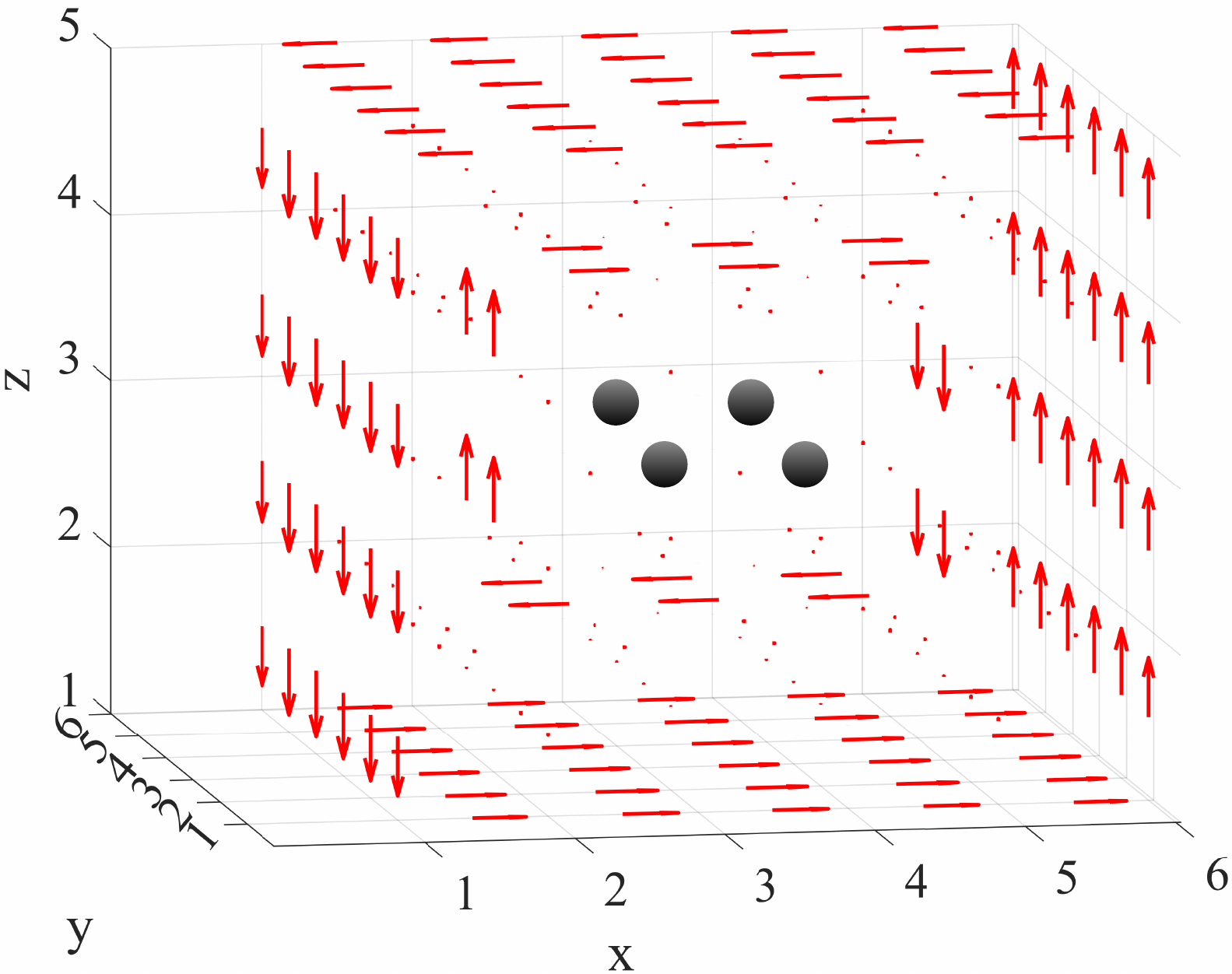}
	\caption{Bosonic current distribution in the presence of bulk defects (gray circles), with magnetic field $\mathbf{b} =(0,0.1,0)$.}
	\label{Fig:volumetric_defect}
\end{figure}
\begin{figure*}[t]
	\centering	\begin{minipage}{0.3\linewidth}\begin{flushleft}(a)\end{flushleft} \centering
		\includegraphics[width=\linewidth]{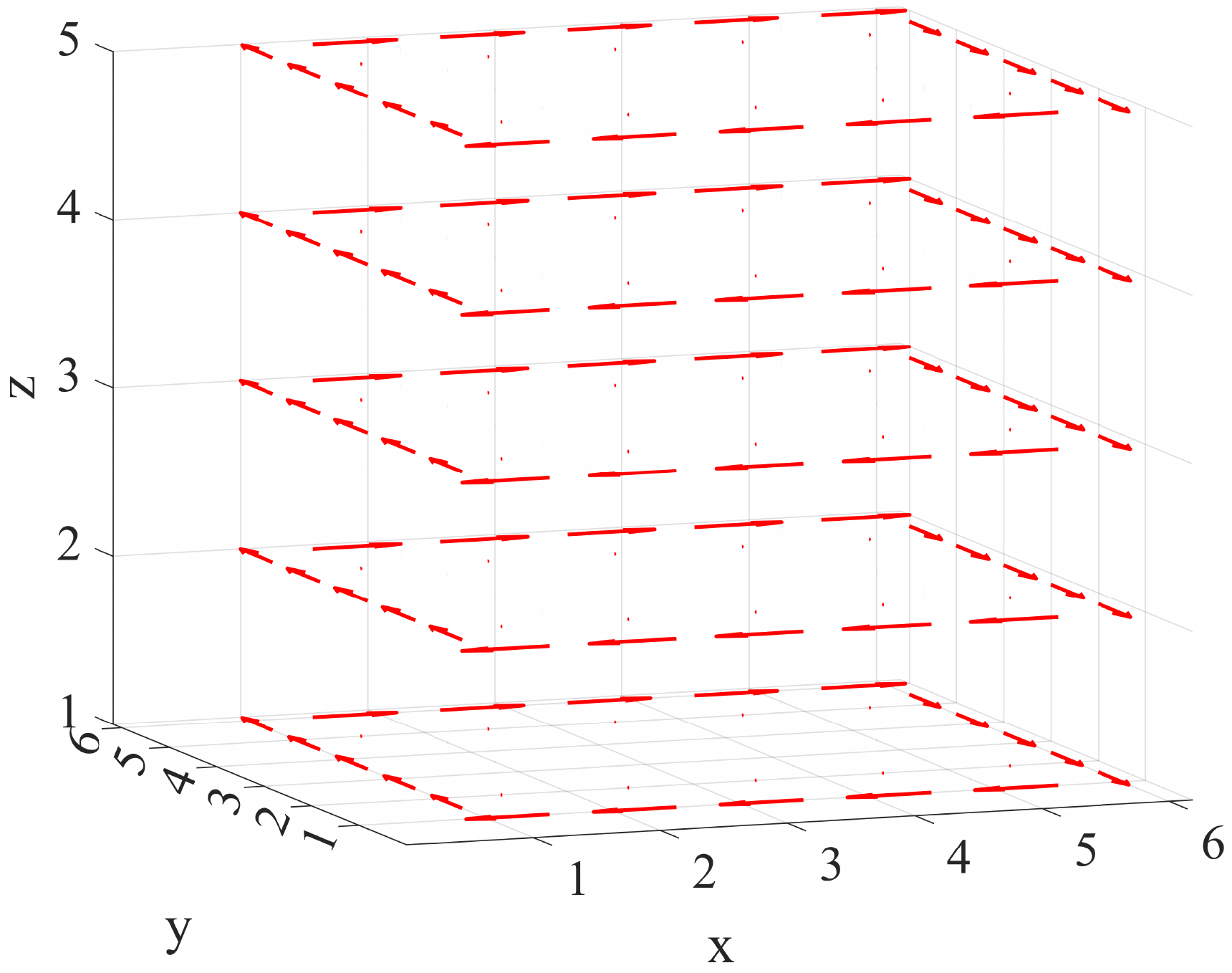}
	\end{minipage}
	\hspace{5.6mm}
	\begin{minipage}{0.31\linewidth}		
		\begin{flushleft}(b)\end{flushleft} 
		\centering
		\includegraphics[width=\linewidth]{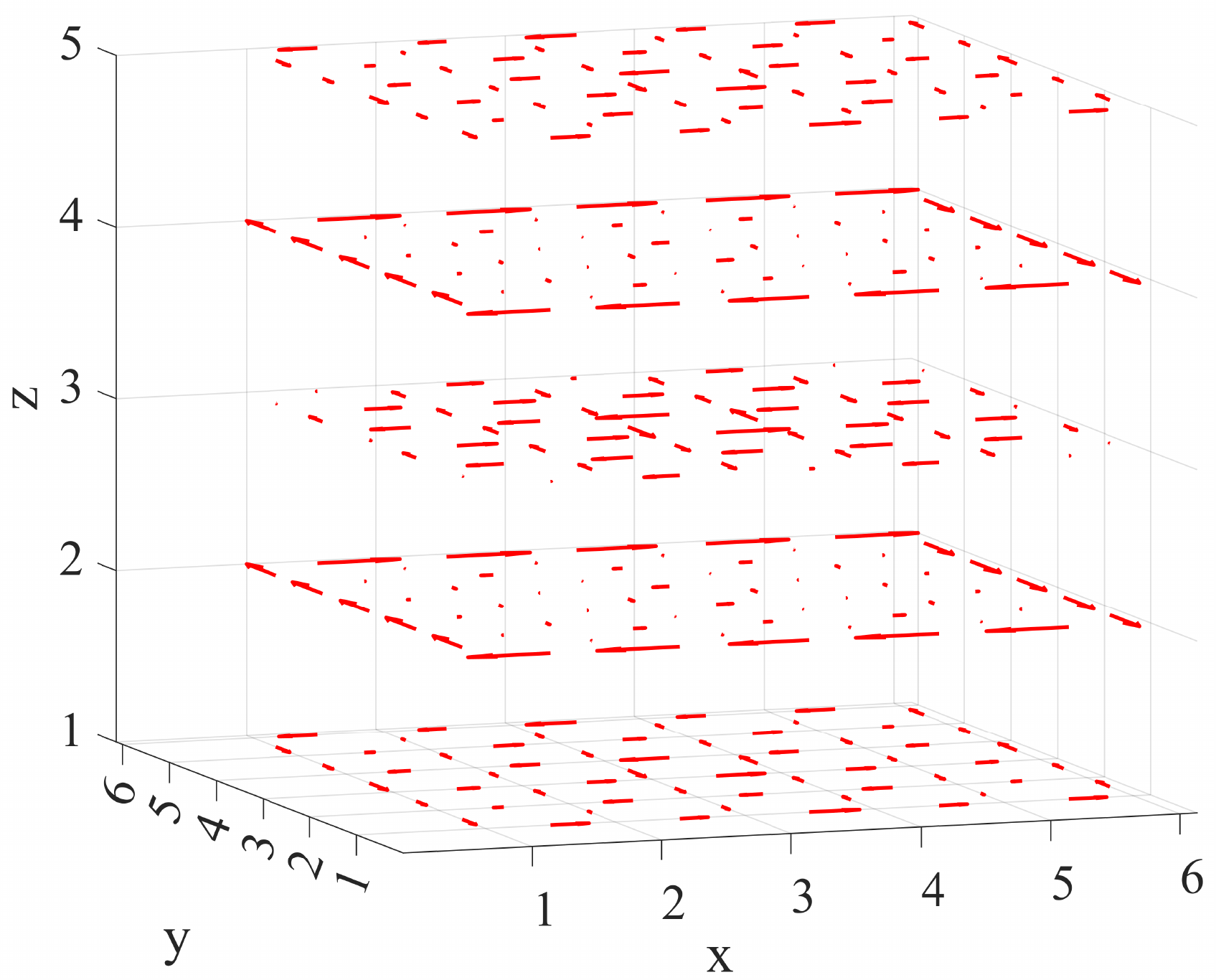}
	\end{minipage}
	\hspace{4mm}
	\begin{minipage}{0.32\linewidth}
		\begin{flushleft}(c)\end{flushleft} 
		\centering
		\includegraphics[width=\linewidth]{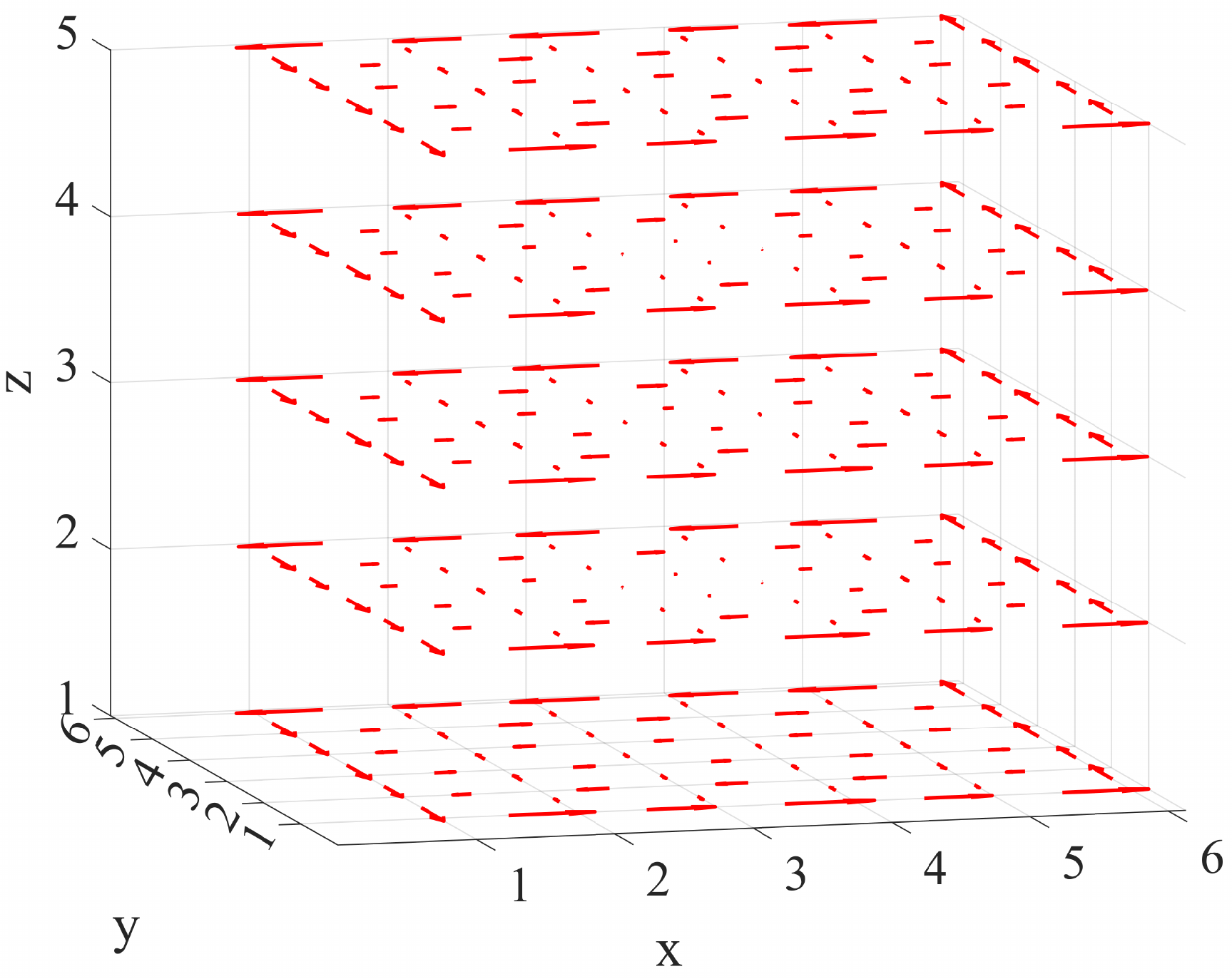}
	\end{minipage}
	
	\caption{Fermionic currents under a vertical magnetic field, $\mathbf{b}=(0,0,0.4)$. (a) Edge currents emerge due to the erasure effect far from half filling, with chemical potential $\mu=\omega_0-4t$.  (b) The edge currents are strongly disrupted near half-filling, $\mu=\omega_0$. (c)  Same as (b) but with weaker tunnelling in the $Z$ direction, $t_Z = 0.1 t$. The currents are now concentrated near the boundary as the system becomes quasi-2D. \label{fig:fermions}}
\end{figure*}

In the previous examples, the protecting symmetries reduce in fact to 2D symmetries in planes, which may lead to stable situations depending on the orientation of $\mathbf{b}$. Therefore, those 2D symmetries are unstable under tilting of $\mathbf{b}$ from directions orthogonal or parallel to those planes. In order to obtain a surface current that is unconditionally symmetry-protected in our 3D lattice, we must employ a genuine 3D symmetry. As discussed in Sec.~\ref{sec:symmetries}, such a symmetry is $\Theta \hat{I}$.  Perturbating defects which comply with this symmetry do not spoil the surface current for any orientation of $\mathbf{b}$ or values of $t_{X,Y,Z}\neq0$.  This is illustrated in Fig.~\ref{fig:defects}(c), where defects placed on opposite corners of the cuboid do not destroy the boundary currents despite the genuinely 3D field configuration, $\mathbf{b} = (0.1,0.1,0.1)$. Instead, the edge currents simply detour around the impurity sites. When the tunnelling in the $Z$ direction is switched off, however, the defects strongly disrupt the current pattern on the planes to which they are confined [Fig.~\ref{fig:defects}(f)].

We note that the point-like character of the defects employed so far does not play any relevant role in the robustness of the currents. Surface and volumentric defects can be introduced leading to stable currents provided that the protection symmetry is satisfied; see Fig.~\ref{Fig:volumetric_defect}, for example. Here, as in Fig.~\ref{fig:defects}, we observe that the introduction of defects in a symmetric configuration leads to the emergence of counter-propagating currents that ``shield'' the impurities. This can be understood as a manifestation of the erasure effect~\cite{Rivas2017,Mitchison2021}, since the defects effectively generate a new boundary within the system.

\subsection{Fermionic lattices}

In the fermionic case, the situation is the same as for bosons at relatively high temperatures $T_{c,h}\gtrsim t_{X,Y,Z}$, where particle exchange statistics plays little role. In contrast, fermions and bosons behave very differently when one of the reservoirs is at very low temperature, as in our examples. In this case, the existence of surface currents depends strongly on the value of the chemical potential, which, via the non-equilibrium distribution function~\eqref{mode_occ_symmetry}, selects the portions of the single-particle spectrum that contribute significantly to the current pattern.  In Fig.~\ref{fig:fermions}(a) we show that, for $\mu\ll \omega_0$, the fermionic system behaves similarly to the bosonic one: the current is fully localised on the surface of the system. These fermionic surface currents enjoy the same robustness against symmetric defects as in the bosonic case. This behaviour is explained by the erasure effect described in Sec.~\ref{sec:symmetries}.

Close to half-filling, however, the situation is quite different as shown in Figs.~\ref{fig:fermions}(b,c). Here, the terminology half-filling refers to the case $\mu=\omega_0$, which would correspond to exactly half-filled bands at zero temperature. If the tunnelling in one direction is significantly weaker, e.g. $t_Z\ll t_{X,Y}$, the system behaves like a set of weakly coupled 2D layers, each hosting edge modes near energy $\omega\approx \omega_0$. These modes are the predominant carrier of current when the system is half-filled, leading to currents localised near the boundaries as shown in Fig.~\ref{fig:fermions}(b)~\cite{Rivas2017}. These currents are robust against any distribution of impurities in the bulk of the system, irrespective of their symmetries~\cite{Mitchison2021}. However, tunnelling between the layers causes hybridisation of the 2D band structure~\cite{Hasegawa3DQHE,KunsztQHE3D} --- apart from at very specific values of the field and tunnelling amplitudes~\cite{KoshinoButterfly3D}. This hybridisation destroys the edge modes and thus fermionic surface currents typically do not appear for isotropic tunnelling when one of the reservoirs is at low temperature [Fig.~\ref{fig:fermions}(c)].

\section{Conclusions}
\label{sec:conclusions}

In summary, we have studied the distribution of currents within a 3D Hofstadter lattice driven far from equilibrium by weakly coupled particle reservoirs at very different temperatures. We have found dissipatively robust surface currents, which are stable against the introduction of defects in configurations that respect certain nonequilibrium symmetries.  In particular, we demonstrated the existence of surface crosscurrents flowing against the temperature gradient within a 2D manifold on the boundary of the system. These results represent the 3D generalization of the robust boundary currents in 2D models reported in Refs.~\cite{Rivas2017,Mitchison2021}.

Notably, however, we find new effects in 3D that cannot be understood in terms of the quasi-2D physics of concatenated layers. Specifically, in the bosonic case, we have found that the presence of tunnelling in all three directions provides additional stability to certain defect configurations that would otherwise destroy the boundary currents [Fig.~\ref{fig:defects}]. We also found strikingly different behaviour for fermions when one reservoir is at low temperature. Here, the existence of three-dimensional tunnelling tends to destroy the edge modes responsible for boundary currents [Fig.~\ref{fig:fermions}]. Remarkably, therefore, in three dimensions these surface effects are more typical of the bosonic system than the fermionic one, in contrast to the usual intuition from topological physics in 2D.

Our results could be experimentally tested in photonic~\cite{Hafezi2013} or cold-atom~\cite{Bloch,Miyake2013} systems, in which synthetic realisations of the Hofstadter Hamiltonian have already been achieved, or even in solid-state systems where the 3D quantum Hall effect was recently observed~\cite{TangQHE3D}. This would further augment the panoply of exotic boundary phenomena that can be explored in higher-dimensional quantum systems out of equilibrium.

\acknowledgements

M.~T.~M. is supported by a Royal Society-Science Foundation Ireland University Research Fellowship, and  acknowledges funding from the ERC Starting Grant ODYSSEY (Grant  Agreement No.~758403) and the Engineering and Physical Sciences Research Council-Science Foundation Ireland Joint Funding of Research project QuamNESS. A.~R. and M.~A.~M.-D. acknowledge financial support from the Spanish MINECO grants MINECO/FEDER Projects FIS2017-91460-EXP, PGC2018-099169-B-I00 FIS-2018 and from CAM/FEDER Project No. S2018/TCS-4342 (QUITEMAD-CM). The research of A.~R. and M.~A.~M.-D. has been partially supported by the U.S. Army Research Office through Grant No.~W911NF-14-1-0103. Calculations were performed on the GICC cluster at UCM and the Lonsdale cluster maintained by the Trinity Centre for High Performance Computing, which was funded through grants from Science Foundation Ireland. 
	
	\bibliographystyle{apsrev4-2}
	\bibliography{bibliography}

\begin{thebibliography}{59}%
\makeatletter
\providecommand \@ifxundefined [1]{%
 \@ifx{#1\undefined}
}%
\providecommand \@ifnum [1]{%
 \ifnum #1\expandafter \@firstoftwo
 \else \expandafter \@secondoftwo
 \fi
}%
\providecommand \@ifx [1]{%
 \ifx #1\expandafter \@firstoftwo
 \else \expandafter \@secondoftwo
 \fi
}%
\providecommand \natexlab [1]{#1}%
\providecommand \enquote  [1]{``#1''}%
\providecommand \bibnamefont  [1]{#1}%
\providecommand \bibfnamefont [1]{#1}%
\providecommand \citenamefont [1]{#1}%
\providecommand \href@noop [0]{\@secondoftwo}%
\providecommand \href [0]{\begingroup \@sanitize@url \@href}%
\providecommand \@href[1]{\@@startlink{#1}\@@href}%
\providecommand \@@href[1]{\endgroup#1\@@endlink}%
\providecommand \@sanitize@url [0]{\catcode `\\12\catcode `\$12\catcode
  `\&12\catcode `\#12\catcode `\^12\catcode `\_12\catcode `\%12\relax}%
\providecommand \@@startlink[1]{}%
\providecommand \@@endlink[0]{}%
\providecommand \url  [0]{\begingroup\@sanitize@url \@url }%
\providecommand \@url [1]{\endgroup\@href {#1}{\urlprefix }}%
\providecommand \urlprefix  [0]{URL }%
\providecommand \Eprint [0]{\href }%
\providecommand \doibase [0]{https://doi.org/}%
\providecommand \selectlanguage [0]{\@gobble}%
\providecommand \bibinfo  [0]{\@secondoftwo}%
\providecommand \bibfield  [0]{\@secondoftwo}%
\providecommand \translation [1]{[#1]}%
\providecommand \BibitemOpen [0]{}%
\providecommand \bibitemStop [0]{}%
\providecommand \bibitemNoStop [0]{.\EOS\space}%
\providecommand \EOS [0]{\spacefactor3000\relax}%
\providecommand \BibitemShut  [1]{\csname bibitem#1\endcsname}%
\let\auto@bib@innerbib\@empty
\bibitem [{\citenamefont {Auerbach}(1994)}]{auerbach1994interacting}%
  \BibitemOpen
  \bibfield  {author} {\bibinfo {author} {\bibfnamefont {A.}~\bibnamefont
  {Auerbach}},\ }\href {https://books.google.es/books?id=jAmbQgAACAAJ} {\emph
  {\bibinfo {title} {Interacting Electrons and Quantum Magnetism}}},\ Graduate
  texts in contemporary physics\ (\bibinfo  {publisher} {Springer-Verlag},\
  \bibinfo {year} {1994})\BibitemShut {NoStop}%
\bibitem [{\citenamefont {Gonzalez}\ \emph {et~al.}(2008)\citenamefont
  {Gonzalez}, \citenamefont {Martin-Delgado}, \citenamefont {Sierra},\ and\
  \citenamefont {Vozmediano}}]{gonzalez2008quantum}%
  \BibitemOpen
  \bibfield  {author} {\bibinfo {author} {\bibfnamefont {J.}~\bibnamefont
  {Gonzalez}}, \bibinfo {author} {\bibfnamefont {M.}~\bibnamefont
  {Martin-Delgado}}, \bibinfo {author} {\bibfnamefont {G.}~\bibnamefont
  {Sierra}},\ and\ \bibinfo {author} {\bibfnamefont {A.}~\bibnamefont
  {Vozmediano}},\ }\href {https://books.google.es/books?id=z7oPBwAAQBAJ} {\emph
  {\bibinfo {title} {Quantum Electron Liquids and High-Tc
  Superconductivity}}},\ Lecture Notes in Physics Monographs\ (\bibinfo
  {publisher} {Springer Berlin Heidelberg},\ \bibinfo {year}
  {2008})\BibitemShut {NoStop}%
\bibitem [{\citenamefont {Ezawa}(2008)}]{ezawa2008quantum}%
  \BibitemOpen
  \bibfield  {author} {\bibinfo {author} {\bibfnamefont {Z.}~\bibnamefont
  {Ezawa}},\ }\href {https://books.google.es/books?id=4A48DQAAQBAJ} {\emph
  {\bibinfo {title} {Quantum Hall Effects: Field Theoretical Approach And
  Related Topics (2nd Edition)}}}\ (\bibinfo  {publisher} {World Scientific
  Publishing Company},\ \bibinfo {year} {2008})\BibitemShut {NoStop}%
\bibitem [{\citenamefont {Beenakker}(1997)}]{Beenakker1997}%
  \BibitemOpen
  \bibfield  {author} {\bibinfo {author} {\bibfnamefont {C.~W.~J.}\
  \bibnamefont {Beenakker}},\ }\href
  {https://doi.org/10.1103/RevModPhys.69.731} {\bibfield  {journal} {\bibinfo
  {journal} {Rev. Mod. Phys.}\ }\textbf {\bibinfo {volume} {69}},\ \bibinfo
  {pages} {731} (\bibinfo {year} {1997})}\BibitemShut {NoStop}%
\bibitem [{\citenamefont {Rammer}(2018)}]{rammer2018quantum}%
  \BibitemOpen
  \bibfield  {author} {\bibinfo {author} {\bibfnamefont {J.}~\bibnamefont
  {Rammer}},\ }\href {https://books.google.es/books?id=FUpaDwAAQBAJ} {\emph
  {\bibinfo {title} {Quantum Transport Theory}}}\ (\bibinfo  {publisher} {CRC
  Press},\ \bibinfo {year} {2018})\BibitemShut {NoStop}%
\bibitem [{\citenamefont {Nazarov}\ and\ \citenamefont
  {Blanter}(2009)}]{nazarov2009quantum}%
  \BibitemOpen
  \bibfield  {author} {\bibinfo {author} {\bibfnamefont {Y.}~\bibnamefont
  {Nazarov}}\ and\ \bibinfo {author} {\bibfnamefont {Y.}~\bibnamefont
  {Blanter}},\ }\href {https://books.google.es/books?id=bjmXJOFmqZIC} {\emph
  {\bibinfo {title} {Quantum Transport: Introduction to Nanoscience}}}\
  (\bibinfo  {publisher} {Cambridge University Press},\ \bibinfo {year}
  {2009})\BibitemShut {NoStop}%
\bibitem [{\citenamefont {Kitaev}(2003)}]{KITAEV2003}%
  \BibitemOpen
  \bibfield  {author} {\bibinfo {author} {\bibfnamefont {A.}~\bibnamefont
  {Kitaev}},\ }\href
  {https://doi.org/https://doi.org/10.1016/S0003-4916(02)00018-0} {\bibfield
  {journal} {\bibinfo  {journal} {Annals of Physics}\ }\textbf {\bibinfo
  {volume} {303}},\ \bibinfo {pages} {2} (\bibinfo {year} {2003})}\BibitemShut
  {NoStop}%
\bibitem [{\citenamefont {Bombin}\ and\ \citenamefont
  {Martin-Delgado}(2006)}]{Bombin2006}%
  \BibitemOpen
  \bibfield  {author} {\bibinfo {author} {\bibfnamefont {H.}~\bibnamefont
  {Bombin}}\ and\ \bibinfo {author} {\bibfnamefont {M.~A.}\ \bibnamefont
  {Martin-Delgado}},\ }\href {https://doi.org/10.1103/PhysRevLett.97.180501}
  {\bibfield  {journal} {\bibinfo  {journal} {Phys. Rev. Lett.}\ }\textbf
  {\bibinfo {volume} {97}},\ \bibinfo {pages} {180501} (\bibinfo {year}
  {2006})}\BibitemShut {NoStop}%
\bibitem [{\citenamefont {Dennis}\ \emph {et~al.}(2002)\citenamefont {Dennis},
  \citenamefont {Kitaev}, \citenamefont {Landahl},\ and\ \citenamefont
  {Preskill}}]{Dennis2002}%
  \BibitemOpen
  \bibfield  {author} {\bibinfo {author} {\bibfnamefont {E.}~\bibnamefont
  {Dennis}}, \bibinfo {author} {\bibfnamefont {A.}~\bibnamefont {Kitaev}},
  \bibinfo {author} {\bibfnamefont {A.}~\bibnamefont {Landahl}},\ and\ \bibinfo
  {author} {\bibfnamefont {J.}~\bibnamefont {Preskill}},\ }\href
  {https://doi.org/10.1063/1.1499754} {\bibfield  {journal} {\bibinfo
  {journal} {Journal of Mathematical Physics}\ }\textbf {\bibinfo {volume}
  {43}},\ \bibinfo {pages} {4452} (\bibinfo {year} {2002})}\BibitemShut
  {NoStop}%
\bibitem [{\citenamefont {Bombin}\ and\ \citenamefont
  {Martin-Delgado}(2007{\natexlab{a}})}]{Bombin2007}%
  \BibitemOpen
  \bibfield  {author} {\bibinfo {author} {\bibfnamefont {H.}~\bibnamefont
  {Bombin}}\ and\ \bibinfo {author} {\bibfnamefont {M.~A.}\ \bibnamefont
  {Martin-Delgado}},\ }\href {https://doi.org/10.1103/PhysRevLett.98.160502}
  {\bibfield  {journal} {\bibinfo  {journal} {Phys. Rev. Lett.}\ }\textbf
  {\bibinfo {volume} {98}},\ \bibinfo {pages} {160502} (\bibinfo {year}
  {2007}{\natexlab{a}})}\BibitemShut {NoStop}%
\bibitem [{\citenamefont {Qi}\ and\ \citenamefont {Zhang}(2011)}]{QiZhang}%
  \BibitemOpen
  \bibfield  {author} {\bibinfo {author} {\bibfnamefont {X.-L.}\ \bibnamefont
  {Qi}}\ and\ \bibinfo {author} {\bibfnamefont {S.-C.}\ \bibnamefont {Zhang}},\
  }\href {https://doi.org/10.1103/RevModPhys.83.1057} {\bibfield  {journal}
  {\bibinfo  {journal} {Rev. Mod. Phys.}\ }\textbf {\bibinfo {volume} {83}},\
  \bibinfo {pages} {1057} (\bibinfo {year} {2011})}\BibitemShut {NoStop}%
\bibitem [{\citenamefont {Hasan}\ and\ \citenamefont {Kane}(2010)}]{Hasan}%
  \BibitemOpen
  \bibfield  {author} {\bibinfo {author} {\bibfnamefont {M.~Z.}\ \bibnamefont
  {Hasan}}\ and\ \bibinfo {author} {\bibfnamefont {C.~L.}\ \bibnamefont
  {Kane}},\ }\href {https://doi.org/10.1103/RevModPhys.82.3045} {\bibfield
  {journal} {\bibinfo  {journal} {Rev. Mod. Phys.}\ }\textbf {\bibinfo {volume}
  {82}},\ \bibinfo {pages} {3045} (\bibinfo {year} {2010})}\BibitemShut
  {NoStop}%
\bibitem [{\citenamefont {Ando}(2013)}]{Ando2013}%
  \BibitemOpen
  \bibfield  {author} {\bibinfo {author} {\bibfnamefont {Y.}~\bibnamefont
  {Ando}},\ }\href {https://doi.org/10.7566/JPSJ.82.102001} {\bibfield
  {journal} {\bibinfo  {journal} {Journal of the Physical Society of Japan}\
  }\textbf {\bibinfo {volume} {82}},\ \bibinfo {pages} {102001} (\bibinfo
  {year} {2013})}\BibitemShut {NoStop}%
\bibitem [{\citenamefont {Bombin}\ and\ \citenamefont
  {Martin-Delgado}(2007{\natexlab{b}})}]{Bombin2007PRB}%
  \BibitemOpen
  \bibfield  {author} {\bibinfo {author} {\bibfnamefont {H.}~\bibnamefont
  {Bombin}}\ and\ \bibinfo {author} {\bibfnamefont {M.~A.}\ \bibnamefont
  {Martin-Delgado}},\ }\href {https://doi.org/10.1103/PhysRevB.75.075103}
  {\bibfield  {journal} {\bibinfo  {journal} {Phys. Rev. B}\ }\textbf {\bibinfo
  {volume} {75}},\ \bibinfo {pages} {075103} (\bibinfo {year}
  {2007}{\natexlab{b}})}\BibitemShut {NoStop}%
\bibitem [{\citenamefont {Qi}\ \emph {et~al.}(2008)\citenamefont {Qi},
  \citenamefont {Hughes},\ and\ \citenamefont {Zhang}}]{Qi2008}%
  \BibitemOpen
  \bibfield  {author} {\bibinfo {author} {\bibfnamefont {X.-L.}\ \bibnamefont
  {Qi}}, \bibinfo {author} {\bibfnamefont {T.~L.}\ \bibnamefont {Hughes}},\
  and\ \bibinfo {author} {\bibfnamefont {S.-C.}\ \bibnamefont {Zhang}},\ }\href
  {https://doi.org/10.1103/PhysRevB.78.195424} {\bibfield  {journal} {\bibinfo
  {journal} {Phys. Rev. B}\ }\textbf {\bibinfo {volume} {78}},\ \bibinfo
  {pages} {195424} (\bibinfo {year} {2008})}\BibitemShut {NoStop}%
\bibitem [{\citenamefont {Diehl}\ \emph {et~al.}(2011)\citenamefont {Diehl},
  \citenamefont {Rico}, \citenamefont {Baranov},\ and\ \citenamefont
  {Zoller}}]{Diehl2011}%
  \BibitemOpen
  \bibfield  {author} {\bibinfo {author} {\bibfnamefont {S.}~\bibnamefont
  {Diehl}}, \bibinfo {author} {\bibfnamefont {E.}~\bibnamefont {Rico}},
  \bibinfo {author} {\bibfnamefont {M.~A.}\ \bibnamefont {Baranov}},\ and\
  \bibinfo {author} {\bibfnamefont {P.}~\bibnamefont {Zoller}},\ }\href
  {https://doi.org/10.1038/nphys2106} {\bibfield  {journal} {\bibinfo
  {journal} {Nature Phys.}\ }\textbf {\bibinfo {volume} {7}},\ \bibinfo {pages}
  {971} (\bibinfo {year} {2011})}\BibitemShut {NoStop}%
\bibitem [{\citenamefont {Budich}\ \emph {et~al.}(2015)\citenamefont {Budich},
  \citenamefont {Zoller},\ and\ \citenamefont {Diehl}}]{Budich2015}%
  \BibitemOpen
  \bibfield  {author} {\bibinfo {author} {\bibfnamefont {J.~C.}\ \bibnamefont
  {Budich}}, \bibinfo {author} {\bibfnamefont {P.}~\bibnamefont {Zoller}},\
  and\ \bibinfo {author} {\bibfnamefont {S.}~\bibnamefont {Diehl}},\ }\href
  {https://doi.org/10.1103/PhysRevA.91.042117} {\bibfield  {journal} {\bibinfo
  {journal} {Phys. Rev. A}\ }\textbf {\bibinfo {volume} {91}},\ \bibinfo
  {pages} {042117} (\bibinfo {year} {2015})}\BibitemShut {NoStop}%
\bibitem [{\citenamefont {Iemini}\ \emph {et~al.}(2016)\citenamefont {Iemini},
  \citenamefont {Rossini}, \citenamefont {Fazio}, \citenamefont {Diehl},\ and\
  \citenamefont {Mazza}}]{Iemini2016}%
  \BibitemOpen
  \bibfield  {author} {\bibinfo {author} {\bibfnamefont {F.}~\bibnamefont
  {Iemini}}, \bibinfo {author} {\bibfnamefont {D.}~\bibnamefont {Rossini}},
  \bibinfo {author} {\bibfnamefont {R.}~\bibnamefont {Fazio}}, \bibinfo
  {author} {\bibfnamefont {S.}~\bibnamefont {Diehl}},\ and\ \bibinfo {author}
  {\bibfnamefont {L.}~\bibnamefont {Mazza}},\ }\href
  {https://doi.org/10.1103/PhysRevB.93.115113} {\bibfield  {journal} {\bibinfo
  {journal} {Phys. Rev. B}\ }\textbf {\bibinfo {volume} {93}},\ \bibinfo
  {pages} {115113} (\bibinfo {year} {2016})}\BibitemShut {NoStop}%
\bibitem [{\citenamefont {Linzner}\ \emph {et~al.}(2016)\citenamefont
  {Linzner}, \citenamefont {Wawer}, \citenamefont {Grusdt},\ and\ \citenamefont
  {Fleischhauer}}]{Linzner2016}%
  \BibitemOpen
  \bibfield  {author} {\bibinfo {author} {\bibfnamefont {D.}~\bibnamefont
  {Linzner}}, \bibinfo {author} {\bibfnamefont {L.}~\bibnamefont {Wawer}},
  \bibinfo {author} {\bibfnamefont {F.}~\bibnamefont {Grusdt}},\ and\ \bibinfo
  {author} {\bibfnamefont {M.}~\bibnamefont {Fleischhauer}},\ }\href
  {https://doi.org/10.1103/PhysRevB.94.201105} {\bibfield  {journal} {\bibinfo
  {journal} {Phys. Rev. B}\ }\textbf {\bibinfo {volume} {94}},\ \bibinfo
  {pages} {201105} (\bibinfo {year} {2016})}\BibitemShut {NoStop}%
\bibitem [{\citenamefont {Rivas}\ and\ \citenamefont
  {Martin-Delgado}(2017)}]{Rivas2017}%
  \BibitemOpen
  \bibfield  {author} {\bibinfo {author} {\bibfnamefont {{\'{A}}.}~\bibnamefont
  {Rivas}}\ and\ \bibinfo {author} {\bibfnamefont {M.~A.}\ \bibnamefont
  {Martin-Delgado}},\ }\href {https://doi.org/10.1038/s41598-017-06722-x}
  {\bibfield  {journal} {\bibinfo  {journal} {Sci. Rep.}\ }\textbf {\bibinfo
  {volume} {7}},\ \bibinfo {pages} {6350} (\bibinfo {year} {2017})}\BibitemShut
  {NoStop}%
\bibitem [{\citenamefont {Kawabata}\ \emph {et~al.}(2019)\citenamefont
  {Kawabata}, \citenamefont {Shiozaki}, \citenamefont {Ueda},\ and\
  \citenamefont {Sato}}]{Kawabata2019}%
  \BibitemOpen
  \bibfield  {author} {\bibinfo {author} {\bibfnamefont {K.}~\bibnamefont
  {Kawabata}}, \bibinfo {author} {\bibfnamefont {K.}~\bibnamefont {Shiozaki}},
  \bibinfo {author} {\bibfnamefont {M.}~\bibnamefont {Ueda}},\ and\ \bibinfo
  {author} {\bibfnamefont {M.}~\bibnamefont {Sato}},\ }\href
  {https://doi.org/10.1103/PhysRevX.9.041015} {\bibfield  {journal} {\bibinfo
  {journal} {Phys. Rev. X}\ }\textbf {\bibinfo {volume} {9}},\ \bibinfo {pages}
  {041015} (\bibinfo {year} {2019})}\BibitemShut {NoStop}%
\bibitem [{\citenamefont {Song}\ \emph {et~al.}(2019)\citenamefont {Song},
  \citenamefont {Yao},\ and\ \citenamefont {Wang}}]{Song2019}%
  \BibitemOpen
  \bibfield  {author} {\bibinfo {author} {\bibfnamefont {F.}~\bibnamefont
  {Song}}, \bibinfo {author} {\bibfnamefont {S.}~\bibnamefont {Yao}},\ and\
  \bibinfo {author} {\bibfnamefont {Z.}~\bibnamefont {Wang}},\ }\href
  {https://doi.org/10.1103/PhysRevLett.123.170401} {\bibfield  {journal}
  {\bibinfo  {journal} {Phys. Rev. Lett.}\ }\textbf {\bibinfo {volume} {123}},\
  \bibinfo {pages} {170401} (\bibinfo {year} {2019})}\BibitemShut {NoStop}%
\bibitem [{\citenamefont {Shavit}\ and\ \citenamefont
  {Goldstein}(2020)}]{Shavit2020}%
  \BibitemOpen
  \bibfield  {author} {\bibinfo {author} {\bibfnamefont {G.}~\bibnamefont
  {Shavit}}\ and\ \bibinfo {author} {\bibfnamefont {M.}~\bibnamefont
  {Goldstein}},\ }\href {https://doi.org/10.1103/PhysRevB.101.125412}
  {\bibfield  {journal} {\bibinfo  {journal} {Phys. Rev. B}\ }\textbf {\bibinfo
  {volume} {101}},\ \bibinfo {pages} {125412} (\bibinfo {year}
  {2020})}\BibitemShut {NoStop}%
\bibitem [{\citenamefont {Gau}\ \emph {et~al.}(2020)\citenamefont {Gau},
  \citenamefont {Egger}, \citenamefont {Zazunov},\ and\ \citenamefont
  {Gefen}}]{Gau2020}%
  \BibitemOpen
  \bibfield  {author} {\bibinfo {author} {\bibfnamefont {M.}~\bibnamefont
  {Gau}}, \bibinfo {author} {\bibfnamefont {R.}~\bibnamefont {Egger}}, \bibinfo
  {author} {\bibfnamefont {A.}~\bibnamefont {Zazunov}},\ and\ \bibinfo {author}
  {\bibfnamefont {Y.}~\bibnamefont {Gefen}},\ }\href
  {https://doi.org/10.1103/PhysRevLett.125.147701} {\bibfield  {journal}
  {\bibinfo  {journal} {Phys. Rev. Lett.}\ }\textbf {\bibinfo {volume} {125}},\
  \bibinfo {pages} {147701} (\bibinfo {year} {2020})}\BibitemShut {NoStop}%
\bibitem [{\citenamefont {Lieu}\ \emph {et~al.}(2020)\citenamefont {Lieu},
  \citenamefont {McGinley},\ and\ \citenamefont {Cooper}}]{Lieu2020}%
  \BibitemOpen
  \bibfield  {author} {\bibinfo {author} {\bibfnamefont {S.}~\bibnamefont
  {Lieu}}, \bibinfo {author} {\bibfnamefont {M.}~\bibnamefont {McGinley}},\
  and\ \bibinfo {author} {\bibfnamefont {N.~R.}\ \bibnamefont {Cooper}},\
  }\href {https://doi.org/10.1103/PhysRevLett.124.040401} {\bibfield  {journal}
  {\bibinfo  {journal} {Phys. Rev. Lett.}\ }\textbf {\bibinfo {volume} {124}},\
  \bibinfo {pages} {040401} (\bibinfo {year} {2020})}\BibitemShut {NoStop}%
\bibitem [{\citenamefont {McGinley}\ and\ \citenamefont
  {Cooper}(2020)}]{McGinley2020}%
  \BibitemOpen
  \bibfield  {author} {\bibinfo {author} {\bibfnamefont {M.}~\bibnamefont
  {McGinley}}\ and\ \bibinfo {author} {\bibfnamefont {N.~R.}\ \bibnamefont
  {Cooper}},\ }\href {https://doi.org/10.1038/s41567-020-0956-z} {\bibfield
  {journal} {\bibinfo  {journal} {Nature Phys.}\ }\textbf {\bibinfo {volume}
  {16}},\ \bibinfo {pages} {1181} (\bibinfo {year} {2020})}\BibitemShut
  {NoStop}%
\bibitem [{\citenamefont {Flynn}\ \emph {et~al.}(2021)\citenamefont {Flynn},
  \citenamefont {Cobanera},\ and\ \citenamefont {Viola}}]{Flynn2021}%
  \BibitemOpen
  \bibfield  {author} {\bibinfo {author} {\bibfnamefont {V.~P.}\ \bibnamefont
  {Flynn}}, \bibinfo {author} {\bibfnamefont {E.}~\bibnamefont {Cobanera}},\
  and\ \bibinfo {author} {\bibfnamefont {L.}~\bibnamefont {Viola}},\ }\href
  {https://doi.org/10.1103/PhysRevLett.127.245701} {\bibfield  {journal}
  {\bibinfo  {journal} {Phys. Rev. Lett.}\ }\textbf {\bibinfo {volume} {127}},\
  \bibinfo {pages} {245701} (\bibinfo {year} {2021})}\BibitemShut {NoStop}%
\bibitem [{\citenamefont {Mitchison}\ \emph {et~al.}(2022)\citenamefont
  {Mitchison}, \citenamefont {Rivas},\ and\ \citenamefont
  {Martin-Delgado}}]{Mitchison2021}%
  \BibitemOpen
  \bibfield  {author} {\bibinfo {author} {\bibfnamefont {M.~T.}\ \bibnamefont
  {Mitchison}}, \bibinfo {author} {\bibfnamefont {A.}~\bibnamefont {Rivas}},\
  and\ \bibinfo {author} {\bibfnamefont {M.~A.}\ \bibnamefont
  {Martin-Delgado}},\ }\href {https://doi.org/10.1103/PhysRevLett.128.120403}
  {\bibfield  {journal} {\bibinfo  {journal} {Phys. Rev. Lett.}\ }\textbf
  {\bibinfo {volume} {128}},\ \bibinfo {pages} {120403} (\bibinfo {year}
  {2022})}\BibitemShut {NoStop}%
\bibitem [{\citenamefont {Guo}\ and\ \citenamefont {Poletti}(2016)}]{Guo2016}%
  \BibitemOpen
  \bibfield  {author} {\bibinfo {author} {\bibfnamefont {C.}~\bibnamefont
  {Guo}}\ and\ \bibinfo {author} {\bibfnamefont {D.}~\bibnamefont {Poletti}},\
  }\href {https://doi.org/10.1103/PhysRevA.94.033610} {\bibfield  {journal}
  {\bibinfo  {journal} {Phys. Rev. A}\ }\textbf {\bibinfo {volume} {94}},\
  \bibinfo {pages} {033610} (\bibinfo {year} {2016})}\BibitemShut {NoStop}%
\bibitem [{\citenamefont {Guo}\ and\ \citenamefont {Poletti}(2017)}]{Guo2017}%
  \BibitemOpen
  \bibfield  {author} {\bibinfo {author} {\bibfnamefont {C.}~\bibnamefont
  {Guo}}\ and\ \bibinfo {author} {\bibfnamefont {D.}~\bibnamefont {Poletti}},\
  }\href {https://doi.org/10.1103/PhysRevB.96.165409} {\bibfield  {journal}
  {\bibinfo  {journal} {Phys. Rev. B}\ }\textbf {\bibinfo {volume} {96}},\
  \bibinfo {pages} {165409} (\bibinfo {year} {2017})}\BibitemShut {NoStop}%
\bibitem [{\citenamefont {Xing}\ \emph {et~al.}(2020)\citenamefont {Xing},
  \citenamefont {Xu}, \citenamefont {Balachandran},\ and\ \citenamefont
  {Poletti}}]{Xing2020}%
  \BibitemOpen
  \bibfield  {author} {\bibinfo {author} {\bibfnamefont {B.}~\bibnamefont
  {Xing}}, \bibinfo {author} {\bibfnamefont {X.}~\bibnamefont {Xu}}, \bibinfo
  {author} {\bibfnamefont {V.}~\bibnamefont {Balachandran}},\ and\ \bibinfo
  {author} {\bibfnamefont {D.}~\bibnamefont {Poletti}},\ }\href
  {https://doi.org/10.1103/PhysRevB.102.245433} {\bibfield  {journal} {\bibinfo
   {journal} {Phys. Rev. B}\ }\textbf {\bibinfo {volume} {102}},\ \bibinfo
  {pages} {245433} (\bibinfo {year} {2020})}\BibitemShut {NoStop}%
\bibitem [{\citenamefont {Fu}\ \emph {et~al.}(2007)\citenamefont {Fu},
  \citenamefont {Kane},\ and\ \citenamefont {Mele}}]{FuKaneMele3D2007}%
  \BibitemOpen
  \bibfield  {author} {\bibinfo {author} {\bibfnamefont {L.}~\bibnamefont
  {Fu}}, \bibinfo {author} {\bibfnamefont {C.~L.}\ \bibnamefont {Kane}},\ and\
  \bibinfo {author} {\bibfnamefont {E.~J.}\ \bibnamefont {Mele}},\ }\href
  {https://doi.org/10.1103/PhysRevLett.98.106803} {\bibfield  {journal}
  {\bibinfo  {journal} {Phys. Rev. Lett.}\ }\textbf {\bibinfo {volume} {98}},\
  \bibinfo {pages} {106803} (\bibinfo {year} {2007})}\BibitemShut {NoStop}%
\bibitem [{\citenamefont {Fu}\ and\ \citenamefont {Kane}(2008)}]{FuKane2008}%
  \BibitemOpen
  \bibfield  {author} {\bibinfo {author} {\bibfnamefont {L.}~\bibnamefont
  {Fu}}\ and\ \bibinfo {author} {\bibfnamefont {C.~L.}\ \bibnamefont {Kane}},\
  }\href {https://doi.org/10.1103/PhysRevLett.100.096407} {\bibfield  {journal}
  {\bibinfo  {journal} {Phys. Rev. Lett.}\ }\textbf {\bibinfo {volume} {100}},\
  \bibinfo {pages} {096407} (\bibinfo {year} {2008})}\BibitemShut {NoStop}%
\bibitem [{\citenamefont {Hasan}\ and\ \citenamefont
  {Moore}(2011)}]{HasanMoore3D}%
  \BibitemOpen
  \bibfield  {author} {\bibinfo {author} {\bibfnamefont {M.~Z.}\ \bibnamefont
  {Hasan}}\ and\ \bibinfo {author} {\bibfnamefont {J.~E.}\ \bibnamefont
  {Moore}},\ }\href {https://doi.org/10.1146/annurev-conmatphys-062910-140432}
  {\bibfield  {journal} {\bibinfo  {journal} {Annual Review of Condensed Matter
  Physics}\ }\textbf {\bibinfo {volume} {2}},\ \bibinfo {pages} {55} (\bibinfo
  {year} {2011})}\BibitemShut {NoStop}%
\bibitem [{\citenamefont {Moore}\ and\ \citenamefont
  {Balents}(2007)}]{Moore3D}%
  \BibitemOpen
  \bibfield  {author} {\bibinfo {author} {\bibfnamefont {J.~E.}\ \bibnamefont
  {Moore}}\ and\ \bibinfo {author} {\bibfnamefont {L.}~\bibnamefont
  {Balents}},\ }\href {https://doi.org/10.1103/PhysRevB.75.121306} {\bibfield
  {journal} {\bibinfo  {journal} {Phys. Rev. B}\ }\textbf {\bibinfo {volume}
  {75}},\ \bibinfo {pages} {121306} (\bibinfo {year} {2007})}\BibitemShut
  {NoStop}%
\bibitem [{\citenamefont {Roy}(2009)}]{Roy3D}%
  \BibitemOpen
  \bibfield  {author} {\bibinfo {author} {\bibfnamefont {R.}~\bibnamefont
  {Roy}},\ }\href {https://doi.org/10.1103/PhysRevB.79.195322} {\bibfield
  {journal} {\bibinfo  {journal} {Phys. Rev. B}\ }\textbf {\bibinfo {volume}
  {79}},\ \bibinfo {pages} {195322} (\bibinfo {year} {2009})}\BibitemShut
  {NoStop}%
\bibitem [{\citenamefont {Hsieh}\ \emph {et~al.}(2008)\citenamefont {Hsieh},
  \citenamefont {Qian}, \citenamefont {Wray}, \citenamefont {Xia},
  \citenamefont {Hor}, \citenamefont {Cava},\ and\ \citenamefont
  {Hasan}}]{Hsieh2008}%
  \BibitemOpen
  \bibfield  {author} {\bibinfo {author} {\bibfnamefont {D.}~\bibnamefont
  {Hsieh}}, \bibinfo {author} {\bibfnamefont {D.}~\bibnamefont {Qian}},
  \bibinfo {author} {\bibfnamefont {L.}~\bibnamefont {Wray}}, \bibinfo {author}
  {\bibfnamefont {Y.}~\bibnamefont {Xia}}, \bibinfo {author} {\bibfnamefont
  {Y.~S.}\ \bibnamefont {Hor}}, \bibinfo {author} {\bibfnamefont {R.~J.}\
  \bibnamefont {Cava}},\ and\ \bibinfo {author} {\bibfnamefont {M.~Z.}\
  \bibnamefont {Hasan}},\ }\href {https://doi.org/10.1038/nature06843}
  {\bibfield  {journal} {\bibinfo  {journal} {Nature}\ }\textbf {\bibinfo
  {volume} {452}},\ \bibinfo {pages} {970} (\bibinfo {year}
  {2008})}\BibitemShut {NoStop}%
\bibitem [{\citenamefont {Ortiz}\ and\ \citenamefont
  {Martin-Delgado}(2016)}]{LOrtiz2016}%
  \BibitemOpen
  \bibfield  {author} {\bibinfo {author} {\bibfnamefont {L.}~\bibnamefont
  {Ortiz}}\ and\ \bibinfo {author} {\bibfnamefont {M.}~\bibnamefont
  {Martin-Delgado}},\ }\href
  {https://doi.org/https://doi.org/10.1016/j.aop.2016.10.008} {\bibfield
  {journal} {\bibinfo  {journal} {Annals of Physics}\ }\textbf {\bibinfo
  {volume} {375}},\ \bibinfo {pages} {193} (\bibinfo {year}
  {2016})}\BibitemShut {NoStop}%
\bibitem [{\citenamefont {Montambaux}\ and\ \citenamefont
  {Kohmoto}(1990)}]{MontambauxQHE3D}%
  \BibitemOpen
  \bibfield  {author} {\bibinfo {author} {\bibfnamefont {G.}~\bibnamefont
  {Montambaux}}\ and\ \bibinfo {author} {\bibfnamefont {M.}~\bibnamefont
  {Kohmoto}},\ }\href {https://doi.org/10.1103/PhysRevB.41.11417} {\bibfield
  {journal} {\bibinfo  {journal} {Phys. Rev. B}\ }\textbf {\bibinfo {volume}
  {41}},\ \bibinfo {pages} {11417} (\bibinfo {year} {1990})}\BibitemShut
  {NoStop}%
\bibitem [{\citenamefont {Hasegawa}(1990)}]{Hasegawa3DQHE}%
  \BibitemOpen
  \bibfield  {author} {\bibinfo {author} {\bibfnamefont {Y.}~\bibnamefont
  {Hasegawa}},\ }\href {https://doi.org/10.1143/JPSJ.59.4384} {\bibfield
  {journal} {\bibinfo  {journal} {Journal of the Physical Society of Japan}\
  }\textbf {\bibinfo {volume} {59}},\ \bibinfo {pages} {4384} (\bibinfo {year}
  {1990})}\BibitemShut {NoStop}%
\bibitem [{\citenamefont {Kunszt}\ and\ \citenamefont
  {Zee}(1991)}]{KunsztQHE3D}%
  \BibitemOpen
  \bibfield  {author} {\bibinfo {author} {\bibfnamefont {Z.}~\bibnamefont
  {Kunszt}}\ and\ \bibinfo {author} {\bibfnamefont {A.}~\bibnamefont {Zee}},\
  }\href {https://doi.org/10.1103/PhysRevB.44.6842} {\bibfield  {journal}
  {\bibinfo  {journal} {Phys. Rev. B}\ }\textbf {\bibinfo {volume} {44}},\
  \bibinfo {pages} {6842} (\bibinfo {year} {1991})}\BibitemShut {NoStop}%
\bibitem [{\citenamefont {Kohmoto}\ \emph {et~al.}(1992)\citenamefont
  {Kohmoto}, \citenamefont {Halperin},\ and\ \citenamefont
  {Wu}}]{KohmotoIntegerQHE3D}%
  \BibitemOpen
  \bibfield  {author} {\bibinfo {author} {\bibfnamefont {M.}~\bibnamefont
  {Kohmoto}}, \bibinfo {author} {\bibfnamefont {B.~I.}\ \bibnamefont
  {Halperin}},\ and\ \bibinfo {author} {\bibfnamefont {Y.-S.}\ \bibnamefont
  {Wu}},\ }\href {https://doi.org/10.1103/PhysRevB.45.13488} {\bibfield
  {journal} {\bibinfo  {journal} {Phys. Rev. B}\ }\textbf {\bibinfo {volume}
  {45}},\ \bibinfo {pages} {13488} (\bibinfo {year} {1992})}\BibitemShut
  {NoStop}%
\bibitem [{\citenamefont {Koshino}\ \emph {et~al.}(2001)\citenamefont
  {Koshino}, \citenamefont {Aoki}, \citenamefont {Kuroki}, \citenamefont
  {Kagoshima},\ and\ \citenamefont {Osada}}]{KoshinoButterfly3D}%
  \BibitemOpen
  \bibfield  {author} {\bibinfo {author} {\bibfnamefont {M.}~\bibnamefont
  {Koshino}}, \bibinfo {author} {\bibfnamefont {H.}~\bibnamefont {Aoki}},
  \bibinfo {author} {\bibfnamefont {K.}~\bibnamefont {Kuroki}}, \bibinfo
  {author} {\bibfnamefont {S.}~\bibnamefont {Kagoshima}},\ and\ \bibinfo
  {author} {\bibfnamefont {T.}~\bibnamefont {Osada}},\ }\href
  {https://doi.org/10.1103/PhysRevLett.86.1062} {\bibfield  {journal} {\bibinfo
   {journal} {Phys. Rev. Lett.}\ }\textbf {\bibinfo {volume} {86}},\ \bibinfo
  {pages} {1062} (\bibinfo {year} {2001})}\BibitemShut {NoStop}%
\bibitem [{\citenamefont {Bernevig}\ \emph {et~al.}(2007)\citenamefont
  {Bernevig}, \citenamefont {Hughes}, \citenamefont {Raghu},\ and\
  \citenamefont {Arovas}}]{BernevigGraphite3D}%
  \BibitemOpen
  \bibfield  {author} {\bibinfo {author} {\bibfnamefont {B.~A.}\ \bibnamefont
  {Bernevig}}, \bibinfo {author} {\bibfnamefont {T.~L.}\ \bibnamefont
  {Hughes}}, \bibinfo {author} {\bibfnamefont {S.}~\bibnamefont {Raghu}},\ and\
  \bibinfo {author} {\bibfnamefont {D.~P.}\ \bibnamefont {Arovas}},\ }\href
  {https://doi.org/10.1103/PhysRevLett.99.146804} {\bibfield  {journal}
  {\bibinfo  {journal} {Phys. Rev. Lett.}\ }\textbf {\bibinfo {volume} {99}},\
  \bibinfo {pages} {146804} (\bibinfo {year} {2007})}\BibitemShut {NoStop}%
\bibitem [{\citenamefont {Tang}\ \emph {et~al.}(2019)\citenamefont {Tang},
  \citenamefont {Ren}, \citenamefont {Wang}, \citenamefont {Zhong},
  \citenamefont {Schneeloch}, \citenamefont {Yang}, \citenamefont {Yang},
  \citenamefont {Lee}, \citenamefont {Gu}, \citenamefont {Qiao},\ and\
  \citenamefont {Zhang}}]{TangQHE3D}%
  \BibitemOpen
  \bibfield  {author} {\bibinfo {author} {\bibfnamefont {F.}~\bibnamefont
  {Tang}}, \bibinfo {author} {\bibfnamefont {Y.}~\bibnamefont {Ren}}, \bibinfo
  {author} {\bibfnamefont {P.}~\bibnamefont {Wang}}, \bibinfo {author}
  {\bibfnamefont {R.}~\bibnamefont {Zhong}}, \bibinfo {author} {\bibfnamefont
  {J.}~\bibnamefont {Schneeloch}}, \bibinfo {author} {\bibfnamefont {S.~A.}\
  \bibnamefont {Yang}}, \bibinfo {author} {\bibfnamefont {K.}~\bibnamefont
  {Yang}}, \bibinfo {author} {\bibfnamefont {P.~A.}\ \bibnamefont {Lee}},
  \bibinfo {author} {\bibfnamefont {G.}~\bibnamefont {Gu}}, \bibinfo {author}
  {\bibfnamefont {Z.}~\bibnamefont {Qiao}},\ and\ \bibinfo {author}
  {\bibfnamefont {L.}~\bibnamefont {Zhang}},\ }\href
  {https://doi.org/10.1038/s41586-019-1180-9} {\bibfield  {journal} {\bibinfo
  {journal} {Nature}\ }\textbf {\bibinfo {volume} {569}},\ \bibinfo {pages}
  {537} (\bibinfo {year} {2019})}\BibitemShut {NoStop}%
\bibitem [{\citenamefont {Moore}\ \emph {et~al.}(2008)\citenamefont {Moore},
  \citenamefont {Ran},\ and\ \citenamefont {Wen}}]{Moore_Hopf}%
  \BibitemOpen
  \bibfield  {author} {\bibinfo {author} {\bibfnamefont {J.~E.}\ \bibnamefont
  {Moore}}, \bibinfo {author} {\bibfnamefont {Y.}~\bibnamefont {Ran}},\ and\
  \bibinfo {author} {\bibfnamefont {X.-G.}\ \bibnamefont {Wen}},\ }\href
  {https://doi.org/10.1103/PhysRevLett.101.186805} {\bibfield  {journal}
  {\bibinfo  {journal} {Phys. Rev. Lett.}\ }\textbf {\bibinfo {volume} {101}},\
  \bibinfo {pages} {186805} (\bibinfo {year} {2008})}\BibitemShut {NoStop}%
\bibitem [{\citenamefont {Deng}\ \emph {et~al.}(2013)\citenamefont {Deng},
  \citenamefont {Wang}, \citenamefont {Shen},\ and\ \citenamefont
  {Duan}}]{Deng_Hopf}%
  \BibitemOpen
  \bibfield  {author} {\bibinfo {author} {\bibfnamefont {D.-L.}\ \bibnamefont
  {Deng}}, \bibinfo {author} {\bibfnamefont {S.-T.}\ \bibnamefont {Wang}},
  \bibinfo {author} {\bibfnamefont {C.}~\bibnamefont {Shen}},\ and\ \bibinfo
  {author} {\bibfnamefont {L.-M.}\ \bibnamefont {Duan}},\ }\href
  {https://doi.org/10.1103/PhysRevB.88.201105} {\bibfield  {journal} {\bibinfo
  {journal} {Phys. Rev. B}\ }\textbf {\bibinfo {volume} {88}},\ \bibinfo
  {pages} {201105} (\bibinfo {year} {2013})}\BibitemShut {NoStop}%
\bibitem [{\citenamefont {Kennedy}(2016)}]{Kennedy_Hopf}%
  \BibitemOpen
  \bibfield  {author} {\bibinfo {author} {\bibfnamefont {R.}~\bibnamefont
  {Kennedy}},\ }\href {https://doi.org/10.1103/PhysRevB.94.035137} {\bibfield
  {journal} {\bibinfo  {journal} {Phys. Rev. B}\ }\textbf {\bibinfo {volume}
  {94}},\ \bibinfo {pages} {035137} (\bibinfo {year} {2016})}\BibitemShut
  {NoStop}%
\bibitem [{\citenamefont {Breuer}\ and\ \citenamefont
  {Petruccione}(2002)}]{Breuer2002}%
  \BibitemOpen
  \bibfield  {author} {\bibinfo {author} {\bibfnamefont {H.-P.}\ \bibnamefont
  {Breuer}}\ and\ \bibinfo {author} {\bibfnamefont {F.}~\bibnamefont
  {Petruccione}},\ }\href
  {https://doi.org/10.1093/acprof:oso/9780199213900.001.0001} {\emph {\bibinfo
  {title} {The theory of open quantum systems}}}\ (\bibinfo  {publisher}
  {Oxford University Press},\ \bibinfo {address} {Oxford},\ \bibinfo {year}
  {2002})\BibitemShut {NoStop}%
\bibitem [{\citenamefont {Rivas}\ and\ \citenamefont
  {Huelga}(2012)}]{RivasHuelga}%
  \BibitemOpen
  \bibfield  {author} {\bibinfo {author} {\bibfnamefont {A.}~\bibnamefont
  {Rivas}}\ and\ \bibinfo {author} {\bibfnamefont {S.~F.}\ \bibnamefont
  {Huelga}},\ }\href {https://doi.org/10.1007/978-3-642-23354-8} {\emph
  {\bibinfo {title} {Open quantum systems. An introduction}}}\ (\bibinfo
  {publisher} {Springer},\ \bibinfo {address} {Heidelberg},\ \bibinfo {year}
  {2012})\BibitemShut {NoStop}%
\bibitem [{\citenamefont {Gardiner}\ and\ \citenamefont
  {Zoller}(2004)}]{GardinerZoller}%
  \BibitemOpen
  \bibfield  {author} {\bibinfo {author} {\bibfnamefont {C.}~\bibnamefont
  {Gardiner}}\ and\ \bibinfo {author} {\bibfnamefont {P.}~\bibnamefont
  {Zoller}},\ }\href {https://www.springer.com/book/9783540223016} {\emph
  {\bibinfo {title} {Quantum noise: a handbook of Markovian and non-Markovian
  quantum stochastic methods with applications to quantum optics}}}\ (\bibinfo
  {publisher} {Springer},\ \bibinfo {address} {Berlin},\ \bibinfo {year}
  {2004})\BibitemShut {NoStop}%
\bibitem [{Note1()}]{Note1}%
  \BibitemOpen
  \bibinfo {note} {Here, the continuity equation is derived by considering the
  Heisenberg equations of motion generated by the microscopic Hamiltonian given
  in Eq.~\protect \textup {\hbox {\mathsurround \z@ \protect \normalfont
  (\ignorespaces \ref {global_Hamiltonian}\unskip \@@italiccorr
  )}}.}\BibitemShut {Stop}%
\bibitem [{\citenamefont {Bu{\v{c}}a}\ and\ \citenamefont
  {Prosen}(2012)}]{Buca_2012}%
  \BibitemOpen
  \bibfield  {author} {\bibinfo {author} {\bibfnamefont {B.}~\bibnamefont
  {Bu{\v{c}}a}}\ and\ \bibinfo {author} {\bibfnamefont {T.}~\bibnamefont
  {Prosen}},\ }\href {https://doi.org/10.1088/1367-2630/14/7/073007} {\bibfield
   {journal} {\bibinfo  {journal} {New J. Phys.}\ }\textbf {\bibinfo {volume}
  {14}},\ \bibinfo {pages} {073007} (\bibinfo {year} {2012})}\BibitemShut
  {NoStop}%
\bibitem [{\citenamefont {Chang}\ and\ \citenamefont {Niu}(1996)}]{Niu}%
  \BibitemOpen
  \bibfield  {author} {\bibinfo {author} {\bibfnamefont {M.-C.}\ \bibnamefont
  {Chang}}\ and\ \bibinfo {author} {\bibfnamefont {Q.}~\bibnamefont {Niu}},\
  }\href {https://doi.org/10.1103/PhysRevB.53.7010} {\bibfield  {journal}
  {\bibinfo  {journal} {Phys. Rev. B}\ }\textbf {\bibinfo {volume} {53}},\
  \bibinfo {pages} {7010} (\bibinfo {year} {1996})}\BibitemShut {NoStop}%
\bibitem [{\citenamefont {Xiao}\ \emph {et~al.}(2010)\citenamefont {Xiao},
  \citenamefont {Chang},\ and\ \citenamefont {Niu}}]{Xiao2010}%
  \BibitemOpen
  \bibfield  {author} {\bibinfo {author} {\bibfnamefont {D.}~\bibnamefont
  {Xiao}}, \bibinfo {author} {\bibfnamefont {M.-C.}\ \bibnamefont {Chang}},\
  and\ \bibinfo {author} {\bibfnamefont {Q.}~\bibnamefont {Niu}},\ }\href
  {https://doi.org/10.1103/RevModPhys.82.1959} {\bibfield  {journal} {\bibinfo
  {journal} {Rev. Mod. Phys.}\ }\textbf {\bibinfo {volume} {82}},\ \bibinfo
  {pages} {1959} (\bibinfo {year} {2010})}\BibitemShut {NoStop}%
\bibitem [{\citenamefont {Price}\ and\ \citenamefont {Cooper}(2012)}]{Cooper}%
  \BibitemOpen
  \bibfield  {author} {\bibinfo {author} {\bibfnamefont {H.~M.}\ \bibnamefont
  {Price}}\ and\ \bibinfo {author} {\bibfnamefont {N.~R.}\ \bibnamefont
  {Cooper}},\ }\href {https://doi.org/10.1103/PhysRevA.85.033620} {\bibfield
  {journal} {\bibinfo  {journal} {Phys. Rev. A}\ }\textbf {\bibinfo {volume}
  {85}},\ \bibinfo {pages} {033620} (\bibinfo {year} {2012})}\BibitemShut
  {NoStop}%
\bibitem [{\citenamefont {Hafezi}\ \emph {et~al.}(2013)\citenamefont {Hafezi},
  \citenamefont {Mittal}, \citenamefont {Fan}, \citenamefont {Migdall},\ and\
  \citenamefont {Taylor}}]{Hafezi2013}%
  \BibitemOpen
  \bibfield  {author} {\bibinfo {author} {\bibfnamefont {M.}~\bibnamefont
  {Hafezi}}, \bibinfo {author} {\bibfnamefont {S.}~\bibnamefont {Mittal}},
  \bibinfo {author} {\bibfnamefont {J.}~\bibnamefont {Fan}}, \bibinfo {author}
  {\bibfnamefont {A.}~\bibnamefont {Migdall}},\ and\ \bibinfo {author}
  {\bibfnamefont {J.~M.}\ \bibnamefont {Taylor}},\ }\href
  {https://doi.org/10.1038/nphoton.2013.274} {\bibfield  {journal} {\bibinfo
  {journal} {Nature Photonics}\ }\textbf {\bibinfo {volume} {7}},\ \bibinfo
  {pages} {1001} (\bibinfo {year} {2013})}\BibitemShut {NoStop}%
\bibitem [{\citenamefont {Aidelsburger}\ \emph {et~al.}(2013)\citenamefont
  {Aidelsburger}, \citenamefont {Atala}, \citenamefont {Lohse}, \citenamefont
  {Barreiro}, \citenamefont {Paredes},\ and\ \citenamefont {Bloch}}]{Bloch}%
  \BibitemOpen
  \bibfield  {author} {\bibinfo {author} {\bibfnamefont {M.}~\bibnamefont
  {Aidelsburger}}, \bibinfo {author} {\bibfnamefont {M.}~\bibnamefont {Atala}},
  \bibinfo {author} {\bibfnamefont {M.}~\bibnamefont {Lohse}}, \bibinfo
  {author} {\bibfnamefont {J.~T.}\ \bibnamefont {Barreiro}}, \bibinfo {author}
  {\bibfnamefont {B.}~\bibnamefont {Paredes}},\ and\ \bibinfo {author}
  {\bibfnamefont {I.}~\bibnamefont {Bloch}},\ }\href
  {https://doi.org/10.1103/PhysRevLett.111.185301} {\bibfield  {journal}
  {\bibinfo  {journal} {Phys. Rev. Lett.}\ }\textbf {\bibinfo {volume} {111}},\
  \bibinfo {pages} {185301} (\bibinfo {year} {2013})}\BibitemShut {NoStop}%
\bibitem [{\citenamefont {Miyake}\ \emph {et~al.}(2013)\citenamefont {Miyake},
  \citenamefont {Siviloglou}, \citenamefont {Kennedy}, \citenamefont {Burton},\
  and\ \citenamefont {Ketterle}}]{Miyake2013}%
  \BibitemOpen
  \bibfield  {author} {\bibinfo {author} {\bibfnamefont {H.}~\bibnamefont
  {Miyake}}, \bibinfo {author} {\bibfnamefont {G.~A.}\ \bibnamefont
  {Siviloglou}}, \bibinfo {author} {\bibfnamefont {C.~J.}\ \bibnamefont
  {Kennedy}}, \bibinfo {author} {\bibfnamefont {W.~C.}\ \bibnamefont
  {Burton}},\ and\ \bibinfo {author} {\bibfnamefont {W.}~\bibnamefont
  {Ketterle}},\ }\href {https://doi.org/10.1103/PhysRevLett.111.185302}
  {\bibfield  {journal} {\bibinfo  {journal} {Phys. Rev. Lett.}\ }\textbf
  {\bibinfo {volume} {111}},\ \bibinfo {pages} {185302} (\bibinfo {year}
  {2013})}\BibitemShut {NoStop}%
\end{thebibliography}%

\clearpage
	
\end{document}